\documentclass[journal]{vgtc} 
\ifpdf%                                % if we use pdflatex
  \pdfoutput=1\relax                   % create PDFs from pdfLaTeX
  \usepackage{graphicx}                % allow us to embed graphics files
  \DeclareGraphicsExtensions{.pdf,.png,.jpg,.jpeg} % for pdflatex we expect .pdf, .png, or .jpg files
\else%                                 % else we use pure latex
  \usepackage{graphicx}                % allow us to embed graphics files
  \DeclareGraphicsExtensions{.eps}     % for pure latex we expect eps files
\fi%

\graphicspath{{figures/}{pictures/}{images/}{./}} % where to search for the images
\usepackage[dvipsnames, svgnames, x11names]{xcolor} 
\usepackage{microtype}                 % use micro-typography (slightly more compact, better to read)
\PassOptionsToPackage{warn}{textcomp}  % to address font issues with \textrightarrow
\usepackage{textcomp}                  % use better special symbols
\usepackage{mathptmx}                  % use matching math font
\usepackage{times}                     % we use Times as the main font
\usepackage{cite}                      % needed to automatically sort the references
\usepackage{tabu}                      % only used for the table example
\usepackage{booktabs} 
\usepackage{bookmark}
\usepackage{amsmath}
\usepackage{stfloats}
\usepackage{lastpage}
\usepackage{fancyhdr}

\usepackage{pifont}
\usepackage{amsmath}
\usepackage{multirow} 
\usepackage{url}
\usepackage{hyperref}  %用于正文引用更加智能一点，自动告诉我们引用的内容是图，表还是公式
\hypersetup{pdfborder={0 0 0} }  %we don't want those silly boxes for links

\usepackage{amsmath}
\usepackage{color}
\usepackage{enumitem}
\usepackage{amsmath}
\usepackage{setspace}
\usepackage{color}
\usepackage{colortbl}
\usepackage{makecell}%导入宏包
\usepackage{array,hhline}
\usepackage{booktabs}
\usepackage[numbers,sort&compress]{natbib}
\usepackage{tikz}
\newcommand*{\circled}[1]{\lower.7ex\hbox{\tikz\draw (0pt, 0pt)%
    circle (.35em) node {\makebox[0.2em][c]{\small #1}};}}
\usepackage[marginal]{footmisc}
\usepackage{bibspacing}
\usepackage[numbers,sort&compress]{natbib}
\usepackage{array}
\usepackage{cite}
\usepackage{setspace}
\onlineid{1069}
\vgtccategory{Research}

\vgtcpapertype{Application}

\title{ASTF: Visual Abstractions of Time-Varying Patterns in Radio Signals}

\author{Ying Zhao, Luhao Ge, Huixuan Xie, Genghuai Bai, Zhao Zhang, Qiang Wei,\\ Yun Lin, Yuchao Liu, and Fangfang Zhou}
\authorfooter{
\item
 Ying Zhao, Luhao Ge, Huixuan Xie, Genghuai Bai, Zhao Zhang, and Fangfang Zhou are with School of Computer Science and Engineering, Central South University, Changsha, China. E-mail: zhaoying@csu.edu.cn, 1014200810@qq.com, 1354168261@qq.com, 1398826924@qq.com, 279966497@qq.com, zff@csu.edu.cn. 
 \item 
 Qiang Wei is with National Key Laboratory of Science and Technology on Blind Signal Processing, Chengdu, China. E-mail: weiqianglg@163.com.
 \item
 Yun Lin is with College of Information and Communication Engineering, Harbin Engineering University, Harbin, China. E-mail: linyun@hrbeu.edu.cn.
 \item
 Yuchao Liu is with China Research Institute of Radiowave Propagation, Qingdao, China. E-mail: lyc8541832@163.com.
 \item
 Fangfang Zhou and Yun Lin are the corresponding authors.
}
\abstract{A time-frequency diagram is a commonly used visualization for observing the time-frequency distribution of radio signals and analyzing their time-varying patterns of communication states in radio monitoring and management. While it excels when performing short-term signal analyses, it becomes inadaptable for long-term signal analyses because it cannot adequately depict signal time-varying patterns in a large time span on a space-limited screen. This research thus presents an abstract signal time-frequency (ASTF) diagram to address this problem. In the diagram design, a visual abstraction method is proposed to visually encode signal communication state changes in time slices. A time segmentation algorithm is proposed to divide a large time span into time slices. Three new quantified metrics and a loss function are defined to ensure the preservation of important time-varying information in the time segmentation. An algorithm performance experiment and a user study are conducted to evaluate the effectiveness of the diagram for long-term signal analyses.}


\begin{document}

\firstsection{Introduction}

\maketitle
%% \section{Introduction} %for journal use above \firstsection{..} instead
% \textsuperscript{\cite{xxx}}实现的是参考文献的上标
A radio signal is an electromagnetic wave used to transmit and receive information through space \cite{A1}. Radio signals vary in their center frequencies and bandwidths because a radio signal must occupy a contiguous section of the electromagnetic spectrum. As the spectrum is a finite natural resource and there has been a rapid growth in radio devices, the demand for the spectrum has continuously increased. Therefore, radio communications must be supervised by radio administration bureaus to ensure orderly spectrum usage \cite{A2,A6}.
\par
\begin{figure}[h]
	\centering
	\vspace{-0.3cm}%调整图片与上文的垂直距离
	\setlength{\abovecaptionskip}{2pt}   %调整图片标题与图距离	
	\setlength{\belowcaptionskip}{-0.5cm}   %调整图片标题与下文距离
	\includegraphics[width=\linewidth]{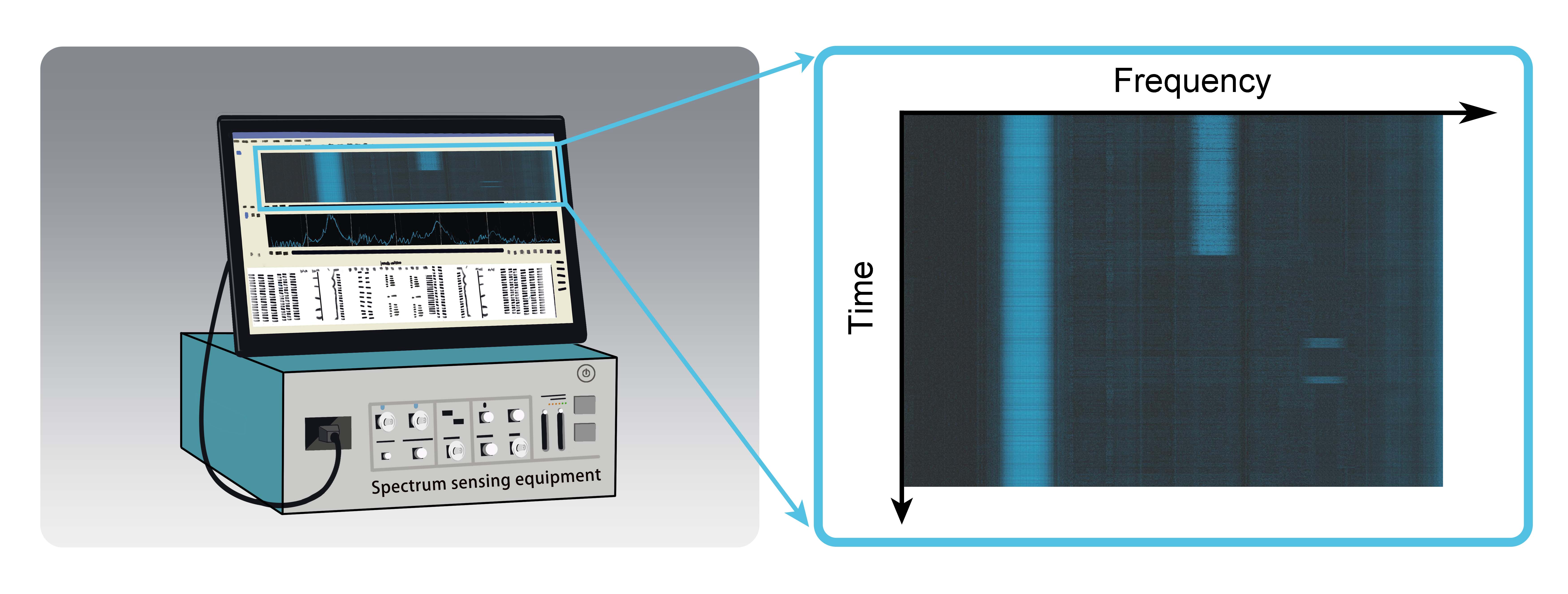}
	\caption{Illustration of a typical TF diagram. The radio spectrum dataset visualized in the diagram covers a 36 MHz frequency band and a period of 3 minutes.}
	\label{fig:figure1}
\end{figure}

{In radio monitoring and management, time-frequency (TF) diagrams are commonly used to visualize radio spectrum data that are captured by spectrum sensing equipment and formatted in frequency frames \cite{A7,A8}. A typical TF diagram is shown in \autoref{fig:figure1}. Its \emph{y}-axis and \emph{x}-axis represent the time and frequency, respectively. A blue band indicates that the electromagnetic amplitudes at the corresponding time-frequency points are higher than environmental noise and may be occupied by a radio signal for communication. For example, through viewing the TF diagram in \autoref{fig:figure1}, three radio signals can be found from left to right. The first signal is constant with high strength. The second signal communicates for nearly one minute. The third signal has low strength and occurs twice for a short time.}

% \footnote{\newline\textit{Manuscript received xx xxx. 201x; accepted xx xxx. 201x. Date of Publication xx xxx. 201x; date of current version xx xxx. 201x. For information on obtaining reprints of this article, please send e-mail to: reprints @ieee.org. Digital Object Identifier:xx.xxxx/TVCG.201x.xxxxxxx}}

\par
\begin{figure*}[ht]
	\centering
	%\vspace{-0.6cm}  %调整图片与上文的垂直距离	
	\setlength{\abovecaptionskip}{2pt}   %调整图片标题与图距离	
	\setlength{\belowcaptionskip}{-0.4cm}   %调整图片标题与下文距离
	\includegraphics[width=\linewidth]{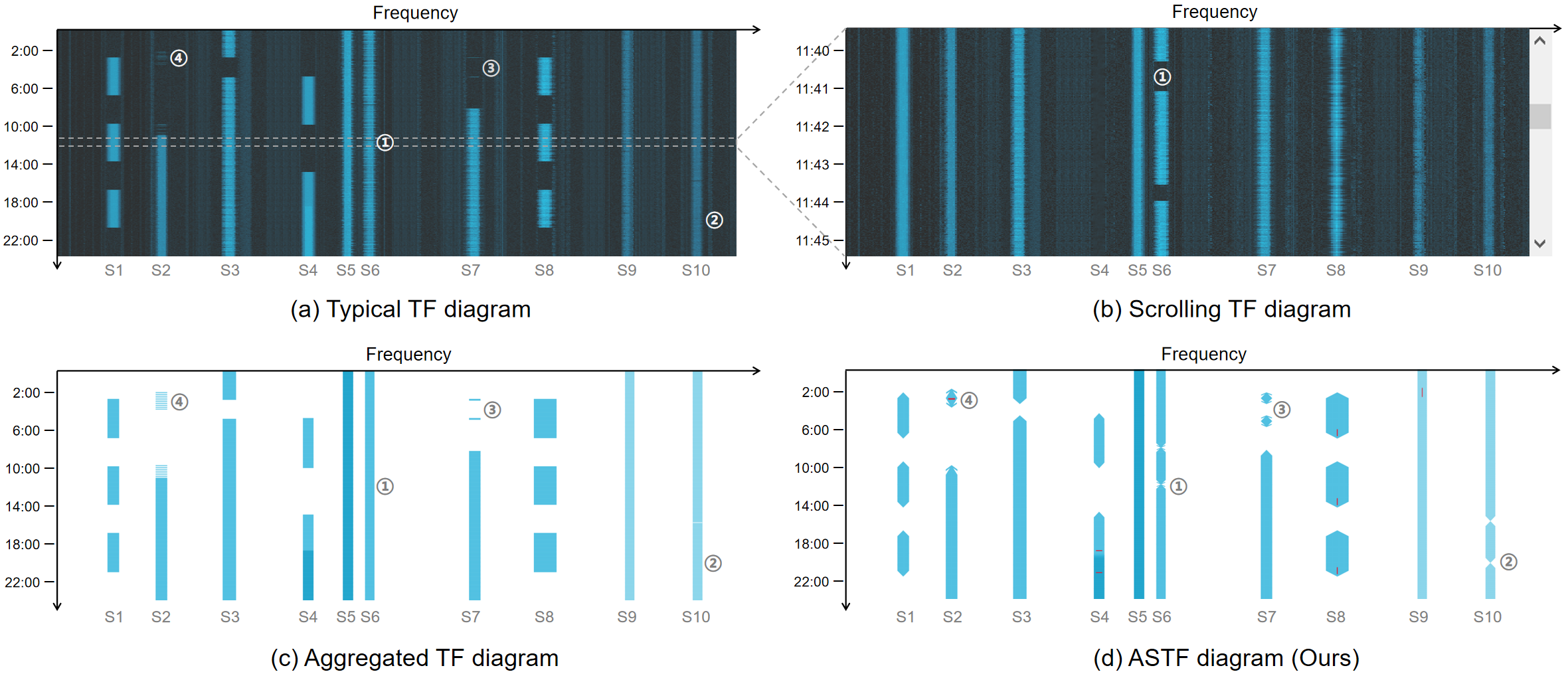}
	\caption{ Illustrations of four visualizations of radio spectrum data with 86,400 frequency frames in 1 day with an interval of 1 second. The display area for visualizing the data has 900 $\times$ 300 pixels. (a) A typical TF diagram visualizes the entire data in the display area, wherein a pixel is rendered 288 times. (b) A scrolling TF diagram visualizes the data with a paging mode using a scrolling bar, wherein 288 pages are filled and each of them contains 300 frequency frames (i.e., 5 minutes). (c) An aggregated TF diagram visualizes the entire data in the display area, wherein the timeline is evenly divided into 300 segments and each of the segments represents the aggregated communication state and strength of a signal in 288 seconds. In this case, a segment will be rendered with color if the proportion of appearing states is larger than 50\% over 288 seconds. Otherwise, the segment remains blank. The color represents the mean strength of a signal over 288 seconds. (d) Our proposed ASTF diagram visualizes the entire data.}
	\label{fig:figure2}
\end{figure*}

{The daily work of radio supervisors includes short-term (e.g., 3$ - $5 minutes) and long-term (e.g., 1$ - $2 days or 1$ - $2 weeks) signal analyses. In short-term analyses, they often rapidly acquire signal details at a specific time or in real-time, such as the number of signals, the center frequency and bandwidth of a signal, or the communication state and strength of a signal. However, in long-term analyses, they are mainly concerned with the changing patterns of communication states and strength, which may indicate a signal’s behavior. TF diagrams excel at performing short-term analyses but become inadaptable for long-term analyses. The main reason is that directly rendering long-term radio spectrum data on a space-limited screen will result in overplotting. Therefore, short discontinuities (i.e., a constant signal disappears for a moment) or short communications (i.e., a signal appears for a short time then disappears) could be not visible, which possibly causes the misjudgment of signal behaviors. Taking the signal S6 in \autoref{fig:figure2} as an example, it communicates for  the entire day, but occurs a few short discontinuities around 11:40 (\autoref{fig:figure2}(b-1)). However, these discontinuities are not visible in \autoref{fig:figure2}(a-1) because overplotting causes many frequency frames to be covered.}

\par{Overplotting \cite{laobaizuishuai} can be partly addressed using animation, scrolling, aggregation, or sampling; however, each of these solutions has challenges. Animation and scrolling cannot directly provide a comprehensive picture over time, so radio supervisors must patiently browse an entire time span to reconstruct the time-varying patterns of signals \cite{A10,A11,A12,A13}, as the scrolling TF diagram illustrated in \autoref{fig:figure2}(b). In addition, aggregation and sampling inevitably cause the loss of information \cite{A14,A15}. Finding a proper aggregation or sampling strategy to preserve short discontinuities and short communications is challenging. For example, the discontinuities of the signal S6 around 11:40 are also not visible in the aggregated TF diagram shown in \autoref{fig:figure2}(c-1), because the duration of each of these discontinuities is less than the threshold of aggregation rendering in this case.}
\par{This research thus presents a new diagram, i.e., \underline{A}bstract \underline{S}ignal \underline{T}ime-\underline{F}requency (ASTF) diagram, that is tailored to long-term radio signal analyses. As shown in \autoref{fig:figure2}(d), the ASTF diagram depicts the time-frequency distribution of 10 signals and their time-varying patterns in 1 day. The ASTF diagram also indicates the occurrences of short discontinuities (\autoref{fig:figure2}(d-1 and d-2)), short communications (\autoref{fig:figure2}(d-3 and d-4)), and signal anomalies (\autoref{fig:figure2}(d-4)). Its design consists of three parts, namely, a data processing (Section \ref{sec:sec4}), a visual abstraction method (Section \ref{sec:sec5}), and a time segmentation algorithm (Section \ref{sec:sec6}).}
\par{During the data processing (\autoref{fig:figure3}(a$ - $d)), radio spectrum data are converted into radio signal data formatted in multidimensional signal records, a set of binarized signal communication state sequences, and a set of signal anomaly records using classic signal processing methods. The radio signal data provide the basic characteristics, namely, center frequency, bandwidth, strength, and signal-to-noise ratio (SNR), of radio signals over time. A binary sequence records the communication states of a signal over time, where logic “1” and logic “0” represent appearing and disappearing states, respectively. Anomalies provide information about abnormal characteristic values of signals, such as sudden frequency shifts or abrupt strength impulses.}
\par{ASTF’s visualization design (\autoref{fig:figure3}(e$ - $h)) adopts the idea of visual abstraction that preserves valuable information while removing meaningless details in data for large-scale data visualizations \cite{A16}. A new visual abstraction method is proposed on the basis of dividing the entire period of a binary sequence into $n$ slices, and the timeline in a display into $n$ cells, to form a one-to-one sequential correspondence. First, ASTF’s basic layout maps signal basic characteristics into a 2D plane with reference to typical TF diagrams. Then, a set of visual encodings for a cell are designed to abstractly represent the communication state changes of a signal in a slice. Lastly, two visual cues are used to highlight anomalies occurring in a signal.}
\par{To ensure that the information needed in long-term signal analyses is well preserved in the visual abstraction, a new time segmentation algorithm, i.e., \underline{B}inary \underline{S}equence \underline{S}egmentation for \underline{V}isual \underline{A}bstraction (BSSVA), is created. Three new quantified metrics are used in the algorithm to measure the preserved information in the visual abstraction. A loss function is also derived to optimize the time segmentation.}
\par{An algorithm performance experiment and a user study were conducted to demonstrate the effectiveness of the BSSVA algorithm and ASTF diagram (Section \ref{sec:sec7}). The performance experiment results showed that the algorithm had fewer information losses than five reference algorithms. The user study results found that participants were more accurate and took less time when using ASTF diagrams to complete long-term radio signal analysis tasks compared to scrolling and aggregated TF diagrams.}
\par{In summary, this paper contributes 1) an ASTF diagram that supports long-term radio signal analyses and 2) a visual abstraction method and a BSSVA time segmentation algorithm that can be applied to long-term binary sequence visualizations.}

%2
\section{Related Work}
\label{sec:sec2}

%2.1
\subsection{Radio Communication Visualizations}
\label{sec:sec2.1}
\vspace{0pt}
Visualizing radio communications is an interdisciplinary domain. Existing studies can be categorized into three well-known visualization subfields, namely, scientific visualization \cite{B1,B2}, information visualization \cite{B3,B4}, and visual analysis \cite{A2,C8}, based on the diversity of data sources and analytical tasks.
\par{Scientific visualization is mainly used for visualizing electromagnetic field simulation data \cite{B2,B8,B9}. Users can, for example, intuitively observe the radiation range of electromagnetic waves of radio devices for performance validations \cite{B10,B12}.}
\par{Information visualization is mainly applied to visualize radio spectrum data. Our approach belongs to this subfield. Classic spectrograms, such as amplitude-time (AT), amplitude-frequency (AF), and TF diagrams, have been widely used in commercial spectrum analyzers \cite{A7,A8}. Researchers have introduced advanced interactions and novel visualizations into these diagrams. Kincaid \cite{C3} used Focus+Context interactions in AT diagrams for analyzing long-lived signals. Hernandez et al. \cite{C4} introduced a time-wrap and 2.5D-layered method into AT diagrams to visualize long-lived signals. While these approaches improve the scalability of AT diagrams in terms of time, they are unsuitable for TF diagrams because radio spectrum data are visualized by lines in AT diagrams but by bitmaps in TF diagrams. Sharakhov et al. \cite{C5} and Cantu et al. \cite{C6} adopted 3D visualizations to present multidimensional information in TF diagrams. Zhao et al. \cite{A2} proposed a new diagram to visualize multiple characteristics of signals. These approaches improve the usability of TF diagrams by enabling them to present multidimensional information but have weak time scalability. ASTF, however, uses a visual abstraction method and a time segmentation algorithm to improve the time scalability of TF diagrams.}
\par{Visual analysis aims to solve complex analytical tasks, such as signal behavior analysis \cite{C8} and electromagnetic situation awareness \cite{A2}. Their main features include multi-source data, advanced data processes, and multi-viewed interactive interfaces. ASTF has the potential to become a common component in such visual analysis systems because it could act as a superior view that provides an overall picture of long-term radio spectrum data.}

\begin{figure*}[ht]
	\centering
	\setlength{\abovecaptionskip}{2pt}   %调整图片标题与图距离	
	\setlength{\belowcaptionskip}{-0.4cm}   %调整图片标题与下文距离
	\includegraphics[width=\linewidth]{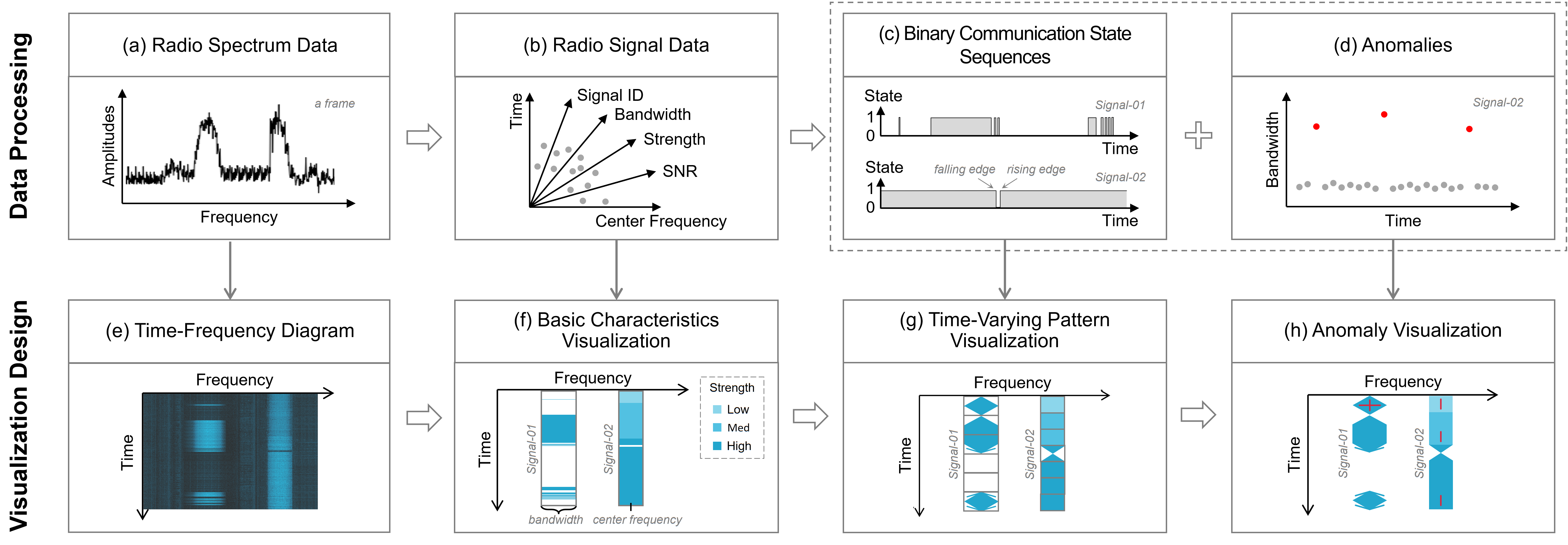}
	\caption{Pipelines of data processing (a$ - $d) and visualization design (e$ - $h). In the data processing pipeline, (a) radio spectrum data formatted in frequency frames are converted into (b) radio signal data formatted in multidimensional data records, then (c) binarized communication state sequences are extracted from and (d) anomalies are detected in the radio signal data. In the visualization pipeline, (e) a TF diagram is commonly used for visualizing short-term radio spectrum data, and an ASTF diagram is newly designed for visualizing long-term radio signal data via three steps: (f) mapping multi-dimensional characteristics of signals into a 2D plane to form the basic layout of ASTF, (g) depicting the time-varying patterns of communication states of signals by using a proposed visual abstraction method and a new time segmentation algorithm, and (h) highlighting anomalies with visual cues.}
	\label{fig:figure3}
\end{figure*}

%2.2
\subsection{Visual Abstractions of Time-Oriented Data}
\label{sec:sec2.2}
Time-oriented data may have a wide range of values and large time spans \cite{A10}, which results in overplotted and cluttered visualizations. Visual abstraction is an important time-oriented data visualization technique that visually encodes important information and removes unnecessary details \cite{A16}. Existing approaches can be divided into three categories by visual channels, namely, position, color, and area \cite{D3,D18}.
\par{Positional visual encodings map positions in a long-time axis into a limited display area using hierarchical or juxtaposed methods. For example, ChronoLenses \cite{D4}, Stack Zooming \cite{D6}, and TimeNotes \cite{D5} layer a large time span with levels of details on demand. Small multiples are a juxtaposed method that visualizes a large time span with juxtaposed sub-spans. Color visual encodings use colors to map a wide range of values into several value levels (e.g., Heatmap \cite{D11,D3}, Horizon Graph \cite{D10}, and Calendar Graph \cite{D17}). Area visual encodings use shapes with different acreages to map a large time span or a wide value range into a visual space (e.g., Icicle plots \cite{Z1,D13}, Sunburst plots \cite{Z2}, and Multi-resolution techniques \cite{D14,D19}). Application systems often combine two or three categories to fulfill specific analytical requirements \cite{D16,D17}. For example, Aigner et al. \cite{D15} proposed a composite visual encoding method based on the three categories to visualize the blood glucose data of diabetes patients. In the present work, the proposed abstraction method is also a coordinated use of these three categories to fulfill long-term radio signal analysis requirements.}

%2.3
\subsection{Time Segmentation Methods}
\label{sec:sec2.3}
In data mining, time segmentation divides a time sequence into appropriate segments for feature engineering \cite{E1,E2}, similarity search \cite{E3,E4}, data compression \cite{E5}, and visual abstraction \cite{E7,D14}. Existing algorithms can be classified into five strategies \cite{E8,E9}. Equal-length division (EL) \cite{E9} divides a sequence into subsequences of equal length. It requires a negligible time cost with a given segment number $n$ but ignores changes and trends in the sequence. Bottom-up (BU) \cite{E8,E11} strategies divide a sequence into small segments and then iteratively merge a pair of adjacent segments until reaching a preset optimization goal. They can obtain an optimal $n$ and segmentation result but incur a high time cost due to their exhaustive searching for an ideal segment pair in every mergence. Top-down (TD) \cite{E8,E11} strategies are natural complements to BUs and recursively split a sequence into two segments until meeting stopping criteria. They obtain an optimal result with a high time cost. Sliding window (SW) \cite{E11,E12} strategies slide a time window from the starting point of a sequence and gradually increase the window size until it satisfies stopping conditions. Then, the stopping point determines a segment and becomes the starting point of the next segment. SWs are ``one-pass" algorithms that have a small time cost but a low optimization level due to their lack of a holistic view. Lastly, feature point-based (FP) \cite{E13,E14} strategies divide a sequence with detected feature points, such as turning or extreme points. They incur a mid-level time cost and are highly effective in reserving changes or trends, but the number of segments is uncertain.
\par{Each of these strategies has advantages and disadvantages. In real-world usage, users should select the proper one or even combine strategies in accordance with target application scenarios. For example, Zhang et al. \cite{E15} used perceptually important points to implement an FP approach to preserve trends in a stock time series. Keogh et al. \cite{E11} used an SW strategy within a BU-based approach to reduce BU’s time cost. The present work combines the EL and FP strategies to create a new time segmentation method for long-term radio signal analyses.}

%3 
\vspace{-0.1cm}
\section{Design Requirements}
\label{sec:sec3}
The target users of this research are radio supervisors who are responsible for supervising radio communications within an area (e.g., an airport or a harbor). Two radio supervisors from a radio spectrum management agency were involved in the entire process of this study. TF diagrams are their main visualization tools for observing and analyzing radio spectrum data in their daily work. They propose several requirements for long-term radio signal analyses, which can be translated into visualization requirements below.
\par{\textbf{R1. Present explicit information about radio signals.} Radio spectrum data record the amplitudes of frequency points but do not directly provide signal characteristics. In short-term analyses, radio supervisors can estimate signal characteristics by observing TF diagrams; however, in long-term analyses, the time span enlarges and the number of signals increases, leading to a considerable cognition burden. Therefore, new diagrams should explicitly present a signal’s characteristics.}
\par{\textbf{R2. Reveal the time-varying patterns of radio signals.}  Radio supervisors have different concerns in short- and long-term analyses. In short-term monitoring, they often acquire signal details. However, in long-term analyses, they are concerned with a signal’s changing patterns of communication states and strength. Therefore, new diagrams should reveal such patterns and ignore unnecessary details.}
\par{\textbf{R3. Highlight anomalies within radio signals.} Anomalies may happen on any signals and often occur over a short time. For example, center frequency shifts may occur when intense noises interfere. Impulse strength peaks may occur when the equipment is unstable. Such anomalies cannot be easily observed in TF diagrams. New diagrams should thus highlight anomalies and depict their distribution over time.}
\par{\textbf{R4. Adopt familiar and easy-to-use visual designs.} As our users are specialized in radio signal analyses, they have relatively solidified cognitive habits to spectrum data visualizations. New diagrams should thus use visual designs that are familiar to these users and be simple and intuitive enough to reduce usage difficulty.}

%4
\section{Data Processing}
\label{sec:sec4}
To fulfill these requirements, radio spectrum data are processed using three steps, i.e., signal identification, communication state inspection, and anomaly detection.

%4.1
\subsection{Signal Identification}
\label{sec:sec4.1}
Signal identification \cite{Z4,Z5} is a historical and evolving field that focuses on extracting explicit signal information from radio spectrum data (R1). In this work, classic signal identification methods recommended by our users are adopted to process radio spectrum data with an interval of 1 second. Compared with semi-structured radio spectrum data, the processed radio signal data are fully structured as multidimensional records (\autoref{fig:figure3}(b)). A record consists of a timestamp, signal ID, and the four basic characteristics of a signal (i.e., center frequency, bandwidth, strength, and SNR). Records with the same signal ID represent the same radio signal.

%4.2
\subsection{Communication State Inspection}
\label{sec:sec4.2}
A radio signal has two communication states, i.e., appearing and disappearing. The two states indicate whether a signal is or is not communicating at a specific time. As radio signal data do not provide this state information, the communication state of a signal at each timestamp needs to be inspected using its characteristics in radio signal data (R2). Based on suggestions from our users, when the strength of a signal in the last 5 seconds is higher than environmental noise, the signal is in the appearing state (i.e., “1”), otherwise it is in the disappearing state (i.e., “0”). Consequently, the communication states of a signal in a period are represented by a binary sequence with an interval of 1 second (\autoref{fig:figure3}(c)).

%4.3
\subsection{Anomaly Detection}
\label{sec:sec4.3}
Anomalies are statistical outliers of signal characteristics, such as frequency shifts and strength impulses. Many existing anomaly detection methods \cite{X1,X2,X3} may be feasible to detect the four types of anomalies (R3) in accordance with the four basic characteristics of a signal. Considering that anomaly detection is not the focus of this paper, we directly adopt classic Pauta criteria \cite{F4,F5} suggested by our users for anomaly detection. For example, the criterion for detecting bandwidth anomalies is that the bandwidth of a signal at a timestamp exceeds the range of ($\mu \pm 3\sigma$), where $\mu$  and $\sigma$  are the average and standard deviation of bandwidth in the period (\autoref{fig:figure3}(d)).

%5
\section{Visualization Design}
\label{sec:sec5}
The visualization design of the ASTF diagram consists of three aspects, i.e., signal characteristics visualization (R1), visual abstractions of signal time-varying patterns (R2), and anomaly visualization (R3).

%5.1
\subsection{Signal Characteristics Visualization}
\label{sec:sec5.1}
The ASTF diagram is designed to visualize radio signal data that provide signal basic characteristics over time. Therefore, the basic characteristic visualization maps the multidimensional information of signals into a 2D plane (R1).

\par{Using the basic layout of TF diagrams and the habits of our users (R4), the \emph{x} and \emph{y} axes are frequency and time, respectively (\autoref{fig:figure3}(f)). A straight stripe along the \emph{y}-axis represents a signal. The stripe’s width and midpoint on the \emph{x}-axis represent the signal’s bandwidth and center frequency, respectively. In general, the center frequency and bandwidth of a signal are constant, and small fluctuations can be ignored in long-term analyses. If large fluctuations occur, they will be detected as anomalies and encoded by visual cues (Section \ref{sec:sec5.3}). The color of a stripe represents the strength. As specific strength values of a signal are unnecessary details during long-term analyses, three strength levels, i.e., high, medium, and low, are used and encoded with blues of different saturations using user-provided thresholds (e.g., -50 dBm and -70 dBm used in this work). These colors are commonly used in spectrum sensing equipment (R4). The SNR is not mapped because it is usually proportional to the strength.}

%5.2
\subsection{Visual Abstractions of Signal Time-Varying Patterns}
\label{sec:sec5.2}
The time-varying patterns of a signal refer to its changing patterns of strength and communication states in a period (R2). The changing patterns of strength are presented using the three-level color encoding method from Section \ref{sec:sec5.1}. This section introduces how to visualize the changing patterns of communication states.
\par{As the communication states of a signal over time are a binary sequence, two types of communication state change points (CSCPs) exist, i.e., rising edge (from disappearing to appearing) and falling edge (from appearing to disappearing), as illustrated in \autoref{fig:figure3}(c). Therefore, the communication states can be represented by a CSCP distribution over time, which contains the number of communications and the starting and ending time of each communication.}
\par{Aggregation is a feasible solution to visualize long-term signal time-varying patterns in a space-limited screen, as illustrated in \autoref{fig:figure2}(c). However, this solution may result in two problems. First, it is difficult to set a proper proportion of appearing states to determine whether a segment should be rendered. A large proportion results in the loss of short communications and a small proportion causes the loss of short discontinuities. For example, Signal S10 has two short discontinuities from 16:00 to 20:00, but only one can be perceived in \autoref{fig:figure2}(c-2). Signal S7 has several short communications from 3:00 to 5:00, but only two short communications are shown in \autoref{fig:figure2}(c-3). Second, if the changes in communication states happen at a certain frequency, several lines or rectangles will be sequentially painted with small gaps (e.g., Signal-01 in \autoref{fig:figure3}(f) and Signal S2 in \autoref{fig:figure2}(c-4)). Such a situation will cause a moiré effect \cite{G1,Z3}, an optical illusion where lines or rectangles align, which will interfere with visual perception.}
\par{To overcome these problems, a visual abstraction method is proposed (\autoref{fig:figure3}(g)). The method consists of three steps.}
\vspace{4pt}
\par{\textbf{STEP 1. Conducting segmentation in data and visual spaces.}}
\vspace{4pt}
\par{A radio signal coexists in data and visual spaces. The data space is the binary communication state sequence. The visual space refers to the stripe representing the signal in an ASTF diagram. This step divides the binary sequence into $n$ time slices and the stripe into $n$ painting cells. The $n$ slices and $n$ cells form a one-to-one sequential correspondence for visual abstraction. Thus, a small cell in visual space conveys the valuable information contained in the corresponding slice that covers a large time span in data space.}
\par{Data space segmentation needs to obtain an optimal $n$ and determine the time span of each slice. The algorithm we propose to handle this is explained in Section \ref{sec:sec6}.}
\par{Visual space segmentation needs to determine the size of a cell and the range of $n$. In terms of cell size, the width of a cell is determined by the bandwidth of the signal. The length of a cell is equal to the division of the total length of the time axis into $n$. That is, all cells in a stripe have the same length to facilitate the subsequent steps. In terms of the range of $n$, we suggest determining it by the size of the painting area in a display from two considerations. First, the search space of optimal $n$ in data space should be limited to a range to reduce the time cost. Second, a large $n$ will generate small-sized cells resulting in low recognizability in visual space, whereas a small $n$ will generate large-sized cells leading to information loss. We conducted pilot studies to obtain an empirical range for mainstreaming displays with a 2K resolution (i.e., [30, 50]).}
\vspace{4pt}
\par{\textbf{STEP 2. Classifying CSCP distributions in data space.}}
\vspace{4pt}
\par{CSCP distributions vary in time slices. If the total number of CSCPs and the type and occurrence time of each CSCP in a slice are considered, visual abstraction would be impossible due to incalculable classes of CSCP distributions. We thus propose a method to classify CSCP distributions into a handful of classes.}
\par
\begin{figure}[h]
	\centering
	%\vspace{-0.6cm}  %调整图片与上文的垂直距离	
	\setlength{\abovecaptionskip}{2pt}   %调整图片标题与图距离	
	\setlength{\belowcaptionskip}{-0.1cm}   %调整图片标题与下文距离
	\includegraphics[width=\linewidth]{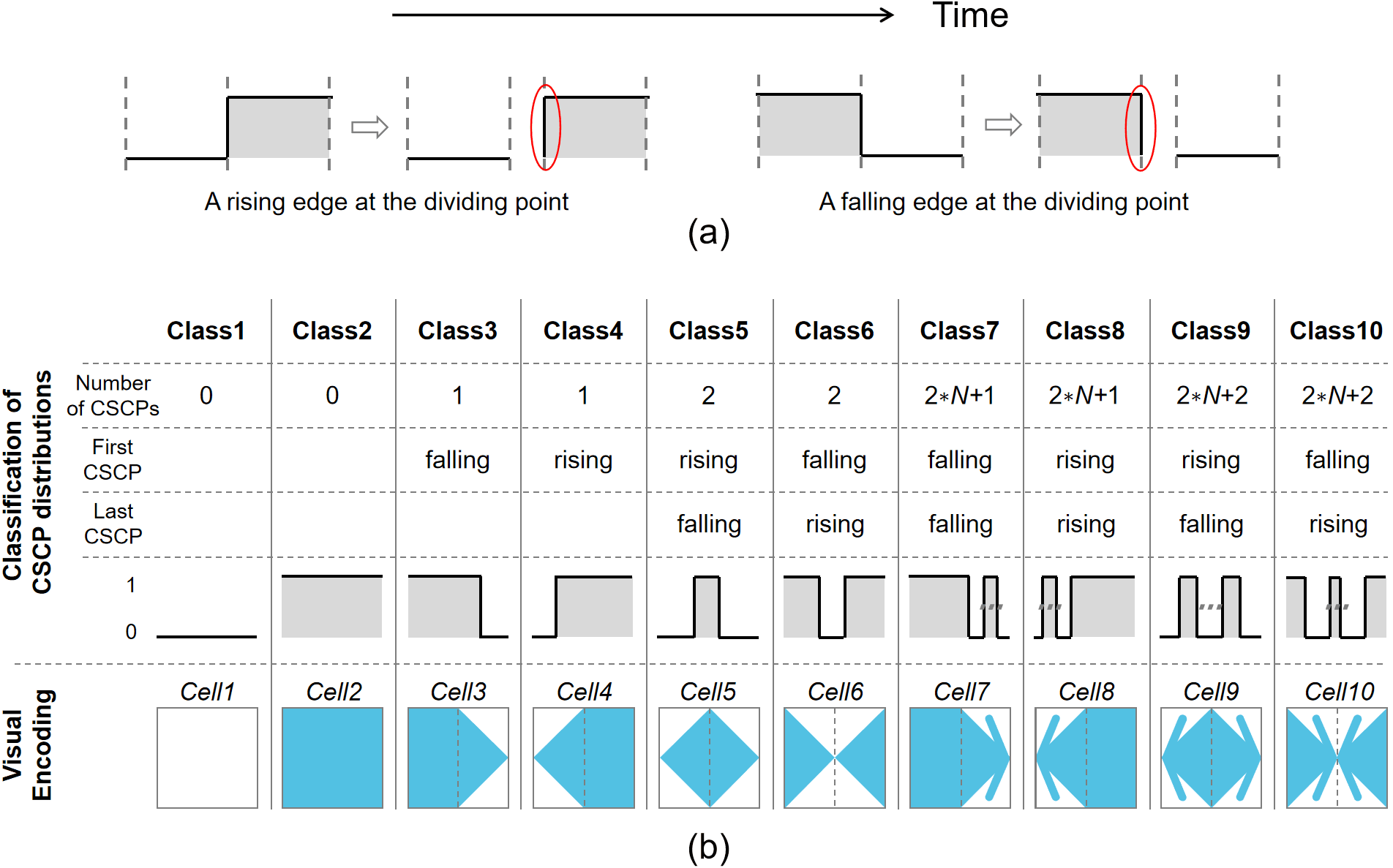}
	\caption{Illustrations of (a) the belonging of a CSCP located on the dividing point of two adjacent slices and (b) classes of CSCP distributions in data space and their corresponding visual encodings in visual space.}
	\label{fig:figure4}
\end{figure}

{The classification method includes an agreement and three bases. The agreement is that if a CSCP occurs at the dividing point of two adjacent slices, it should belong to the slice where the signal appears (\autoref{fig:figure4}(a)). This ensures that a CSCP is only included in one slice. The three bases are the number level of CSCPs, the type of the first CSCP, and the type of the last CSCP in a slice. Using this classification method, ten classes of CSCP distributions are obtained (\autoref{fig:figure4}(b)). Class1 and Class2 have no CSCP, indicating that a signal keeps the same state. Class3 and Class4 have only one CSCP, indicating that a signal changes its state once. Class5 and Class6 have two CSCPs, indicating two state changes. Class5 is a rising-first-falling-second pattern, while Class6 is a falling-first-rising-second pattern. The four remaining classes have more than two CSCPs, presenting falling-first-falling-last (Class7), rising-first-rising-last (Class8), rising-first-falling-last (Class9), and falling-first-rising-last (Class10) patterns.}

\par{This classification method preserves the valuable information concerned by users while discarding unnecessary details that are trivial in long-term signal analyses. Specifically, it preserves the number level of CSCPs in a slice while discarding the specific number. It also preserves the relative occurrence time by the location of a slice in the entire period. For a slice containing two or more CSCPs, it preserves the first and last CSCPs while discarding the other CSCPs, because the two CSCPs indicate a signal’s states when entering and leaving the slice.}
\vspace{4pt}
\par{\textbf{STEP 3. Encoding CSCP distributions in visual space.}}
\vspace{4pt}
\par{This step designs visual encodings to represent the information contained in a CSCP distribution. Four visual encoding rules (VERules) are used to obtain the visual encodings of the 10 classes of CSCP distributions (\autoref{fig:figure4}(b)).}
\par{\emph{VERule1:} For a slice without a CSCP, the cell is (Cell2) or is not (Cell1) filled depending on whether the signal is appearing (Class2) or disappearing (Class1) in the slice.}
\par{\emph{VERule2:} For a slice with only one CSCP, the CSCP type may be a falling or rising edge. A falling edge (Class3) conveys two messages: the signal is communicating when entering the slice, but the communication ends in this slice. We divide the cell (Cell3) into two sides to encode these two messages. The left is filled with a solid rectangle to represent the appearing state of the signal when entering the slice. The right uses a solid isosceles triangle, whose apex is along the direction of the time axis, to represent the occurrence of a falling edge and the disappearing state of the signal when leaving the slice. Likewise, a rising edge (Class4) conveys two messages: the signal is not communicating when entering this slice, but the signal starts a new communication within this slice and is on the appearing state when leaving this slice. We also equally divide the cell (Cell4) into two sides. The left uses a solid isosceles triangle, whose apex is along the opposite direction of the time axis, to represent the occurrence of a rising edge and the disappearing state of the signal when entering the slice. The right is filled with a solid rectangle to represent the appearing state of the signal when leaving the slice.}
\par{\emph{VERule3:} For a slice with two CSCPs, Class5 is a rising-first-falling-second pattern that conveys two messages: the signal is not communicating when entering and leaving the slice but completes a short communication within this slice. We also use two sides of the cell (Cell5) to encode the two messages. The left uses a solid isosceles triangle, whose apex is along the opposite direction of the time axis, to represent the occurrence of a rising edge and the signal’s disappearing state when entering the slice. The right uses another solid isosceles triangle, whose apex is along the direction of the time axis, to represent the occurrence of a falling edge and the signal’s disappearing state when leaving the slice. Class6 is a falling-first-rising-second pattern indicating that the signal is communicating when entering and leaving the slice but has a discontinuity within this slice. We use two solid isosceles triangles with apexes having opposite directions (Cell6) to represent the information conveyed by this pattern.}
\par{\emph{VERule4:} For a slice with more than two CSCPs, four classes of CSCP distributions (i.e., Classes 7$ - $10) exist. We construct a paired relationship between the CSCP distributions of Classes 3$ - $6 and Classes 7$ - $10 in order. In each pair, the signal states when entering and leaving the slice are the same. Therefore, we can directly apply \emph{VERule2} and \emph{VERule3} to Classes 7$ - $10. Moreover, we add two slope lines on the apex of an isosceles triangle (i.e., umbrella-triangle) to represent multiple CSCPs occurring in a slice (e.g., Cells 7$ - $10 in \autoref{fig:figure4}(b)).}
\par
\begin{figure}[h]
	\centering
	%\vspace{-0.6cm}  %调整图片与上文的垂直距离	
	\setlength{\abovecaptionskip}{2pt}   %调整图片标题与图距离	
	\setlength{\belowcaptionskip}{-0.4cm}   %调整图片标题与下文距离
	\includegraphics[width=\linewidth]{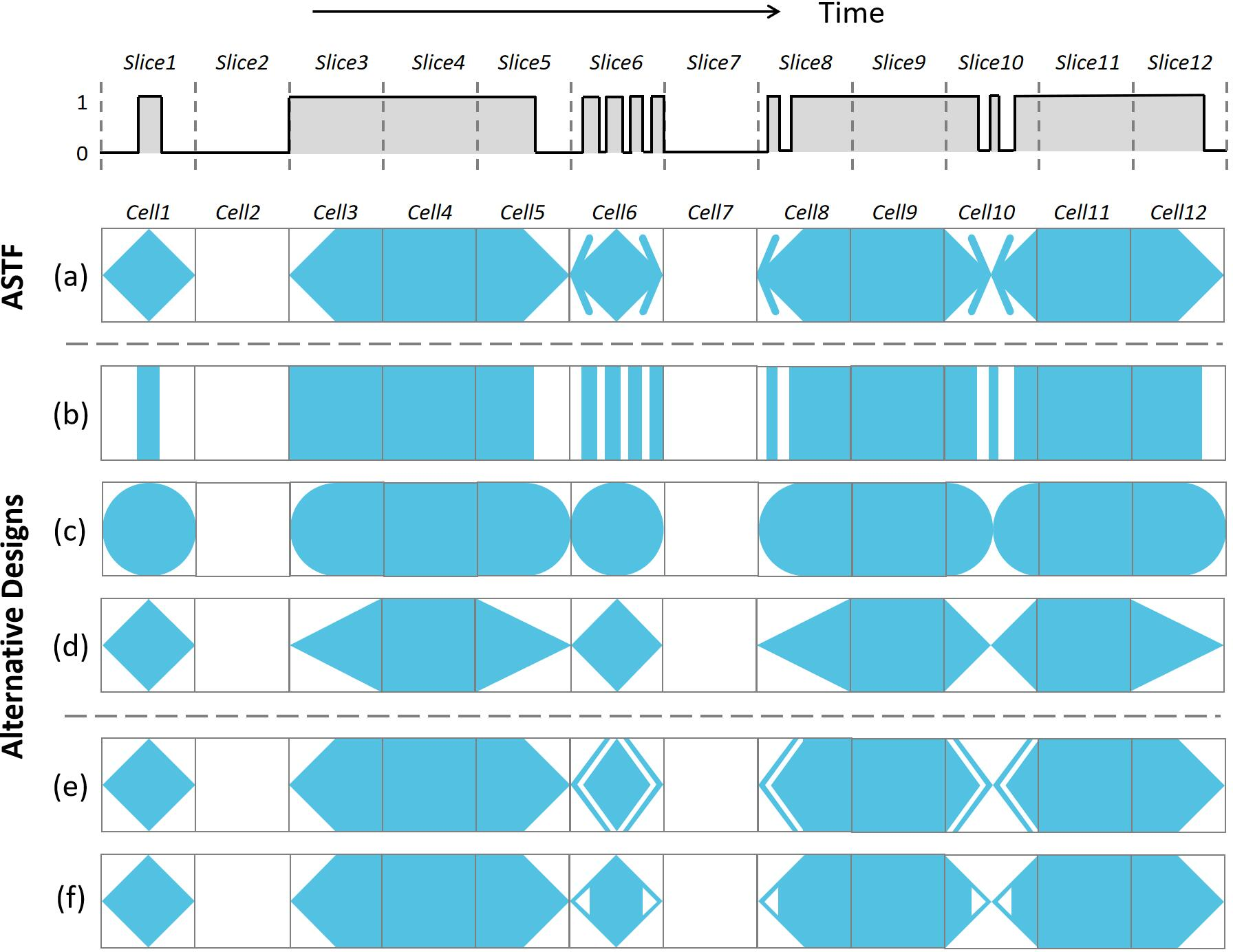}
	\caption{Demonstrations of visual abstraction results of a binary communication state sequence by using (a) the proposed visual encoding method and (b$ - $f) some alternative designs.}
	\label{fig:figure5}
\end{figure}

{This visual encoding method has many advantages. First, the combination of a triangle and rectangle ensures the consistency of communication states of adjacent slices (e.g., Cells 3, 4, and 5 in \autoref{fig:figure5}(a)). Second, isosceles triangles have good visual saliency and stability because angles and blank areas in a cell are visually notable, and the two legs of an angle have the same length. Third, triangles have good semantic clarity because the apex direction of a triangle reflects a CSCP’s type (e.g., a rising edge in Cell3 or a falling edge in Cell5 in \autoref{fig:figure5}(a)), and the apex directions of two triangles in the same cell indicate the occurrence of a single short communication (e.g., Cell1 in \autoref{fig:figure5}(a)) or a single short discontinuity (e.g., Cell6 in \autoref{fig:figure4}(b)) in a slice. Lastly, the frequent occurrences of multiple short communications or short discontinuities in a slice are visually encoded by an umbrella-triangle without a moiré effect (e.g., Cells 6 and 10 in \autoref{fig:figure5}(a)).}
\par{Before arriving at this design, we discussed many alternatives with our target users. For example, using an aggregated TF diagram (e.g., \autoref{fig:figure2}(c) and \autoref{fig:figure5}(b)). We also tried replacing a triangle with a semicircle (\autoref{fig:figure5}(c)); however, the visual stability of semicircles was worse than that of triangles, which could affect viewers’ estimation of CSCP locations on the time axis. We also modified \emph{VERule2} to use one triangle in one cell only. This generated two kinds of triangles with different sizes and angles, which could confuse viewers (e.g., Cells 1 and 3 in \autoref{fig:figure5}(d)). Alternative designs that represented multiple CSCPs occurring in a slice were tested. As two alternatives shown in \autoref{fig:figure5}(e$ - $f), the former caused a moiré effect \cite{G1,Z3}, while the latter was easily confused with the design of the anomaly visualization.}

%5.3
\subsection{Signal Anomaly Visualization}
\label{sec:sec5.3}
To highlight the four types of anomalies (R3), the ASTF diagram supports two visual cues that can be added to a cell. A short red line along the time direction at the center of a cell represents the existence of frequency or/and bandwidth anomalies (\autoref{fig:figure3}(h)). A short red line along the frequency axis direction at the center of a cell represents the existence of strength or/and SNR anomalies. For example, in the first cell of Signal-02 in \autoref{fig:figure3}(h), a  vertical red line represents that frequency or/and bandwidth anomalies occur in the corresponding slice; in the first cell of Signal-01 in \autoref{fig:figure3}(h), vertical and horizontal red lines represent the possible co-occurrence of four types of anomalies.
\par{These visual cues are simple and visually orthogonal to prevent visual interference with each other. The four types of anomalies are grouped to reduce the complexity of visual cues. Empirically, frequency and bandwidth anomalies often co-occur, like the co-occurrence of strength and SNR anomalies. Our users prefer a red line along the time axis to represent frequency or/and bandwidth anomalies (R4).}

%6
\section{Time Segmentation Algorithm}
\label{sec:sec6}
To segment a binary communication state sequence into time slices while fulfilling an optimization goal, a new time segmentation algorithm is proposed. The resulting slices are applied to generate an abstracted time visualization (Section \ref{sec:sec5.2}). This section details the optimization goal definition and the new algorithm.

%6.1
\subsection{Optimization Goal Definition}
\label{sec:sec6.1}
Time segmentation is an optimization problem that depends on a predefined optimization goal. The CSCP distribution in a binary sequence is the main information of interest to users, and the time segmentation is executed in data space while the result is visually encoded in visual space. Therefore, the optimization goal is to minimize the information loss of CSCP distribution during the segmentation and visual encoding to ensure that the visual CSCP distribution in visual space is similar to the actual CSCP distribution in data space. To measure the information loss, three quantified metrics are designed by considering the inner relationships between two spaces.
\vspace{5pt}
\par{\textbf{(1) The similarity of CSCP distributions (Sim\_CD)}}
\vspace{5pt}
\par{Sim\_CD measures the similarity of CSCP distributions between data and visual spaces. Its calculation needs to obtain the exact location of each CSCP on the time axis. However, in visual space, because CSCPs have been encoded by isosceles triangles, there are no exact locations. We solve this problem from both spaces.}
\par
\begin{figure}[h]
	\centering
	\vspace{0.05cm}  %调整图片与上文的垂直距离	
	\setlength{\abovecaptionskip}{2pt}   %调整图片标题与图距离	
	\setlength{\belowcaptionskip}{-0.2cm}   %调整图片标题与下文距离
	\includegraphics[width=\linewidth]{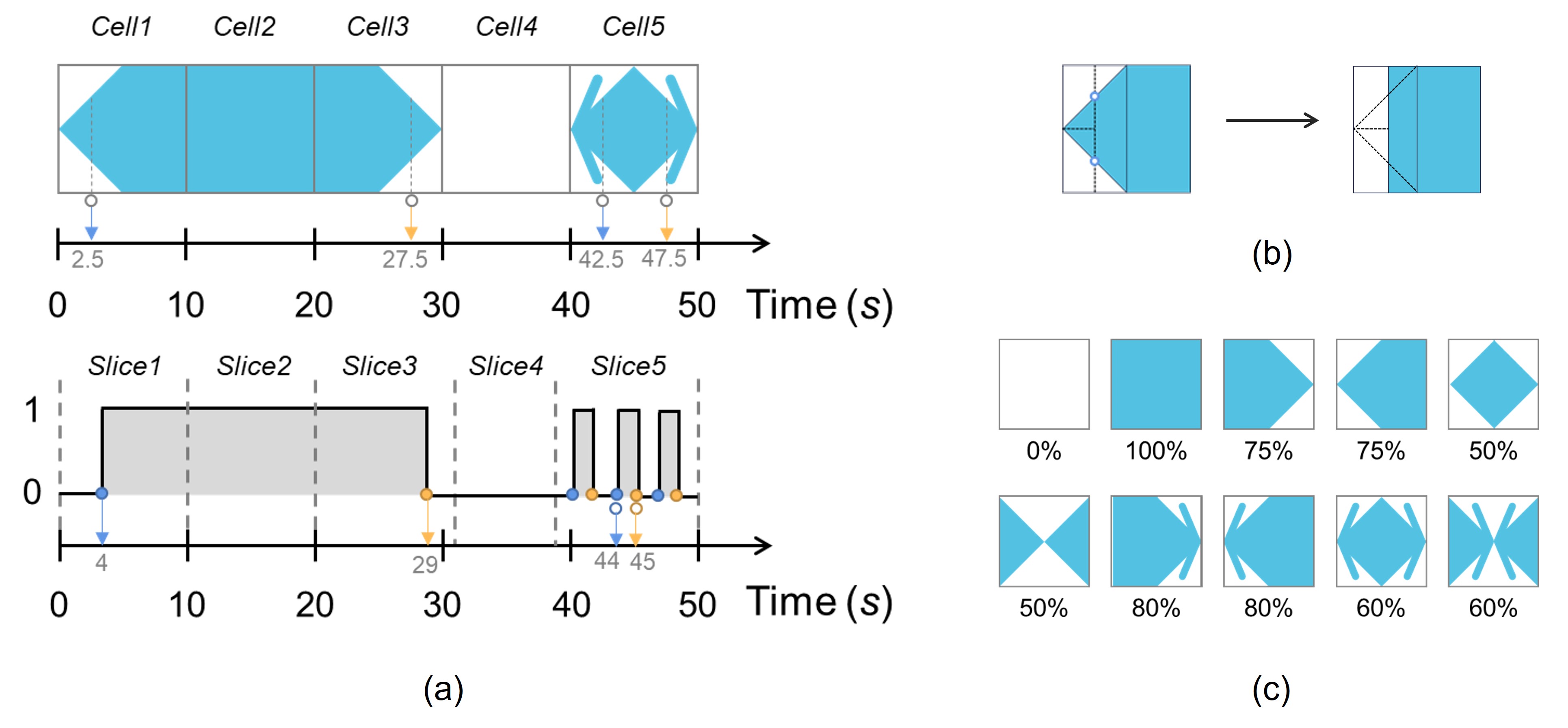}
	\caption{Illustrations of (a) visual CSCP locations, (b) area equivalence principle on a rectangle and an isosceles triangle, and (c) visual duty ratios of 10 classes of CSCP distributions.}
	\label{fig:figure6}
\end{figure}

{In visual space, we stipulate that the point of projecting the midpoints of the legs of an isosceles triangle onto the time axis is the visual location of CSCP(s), regardless of whether this triangle represents one or multiple CSCPs. For example, the visual CSCP location represented by the triangle in Cell1 (\autoref{fig:figure6}(a)) is at 2.5 seconds. This stipulation adopts the principle of area equivalence to estimate a visual CSCP location because area perception is familiar to users and a rectangle that has the same area as an isosceles triangle is easy to find (\autoref{fig:figure6}(b)).}
\par{Thus, using this method, a cell has at most two visual CSCP locations on the time axis. However, in data space, a slice may have more than two CSCPs, and each of which has an exact location. We thus construct a paired relationship between CSCP locations in a slice and in a cell. First, we divide CSCPs into rising and falling groups in a slice. Second, we compute the average of the exact locations of all CSCPs in each group. Thus, a slice can generate, at most, two aggregated CSCP locations. Lastly, the rising and falling groups of CSCPs in a slice correspond to the triangles whose apexes are along the backward and forward directions of the time axis in the corresponding cell in visual space, respectively. In \autoref{fig:figure6}(a), Cell3 has one visual CSCP location (at 27.5 seconds) corresponding to the exact location (at 29 seconds) of the single CSCP in Slice3. Cell5 has two visual CSCP locations. The left (at 42.5 seconds) corresponds to the aggregated CSCP location (at 44 seconds) grouped by the three rising edges in Slice5, and the right (at 47.5 seconds) corresponds to the aggregated CSCP location (at 45 seconds) grouped by the three falling edges in Slice5. Consequently, each visual CSCP location in visual space has its paired aggregated or exact CSCP location in data space.}
\par{After determining the paired relationships of CSCP locations in data and visual spaces, Sim\_CD can thus be formulated as:}
\begin{equation}
	\text{Sim\_CD} = \sum\nolimits_{i = 1}^m {\left| {\frac{{{D_i}}}{T} - \frac{{{V_i}}}{L}} \right|},
	\nonumber
\end{equation}
where $m$ is the number of CSCP location pairs, $V_i$ is the $i$th visual CSCP location in visual space and $D_i$ is the corresponding exact or aggregated CSCP location in data space, $T$ is the entire time span of data space, and $L$ is the entire length of the time axis in visual space. A value approaching 0 indicates a “perfect” result.
\vspace{5pt}
\par{\textbf{(2) The difference in duty ratios (Dif\_DR)}}
\vspace{5pt}
\par{A duty ratio is the proportion of time during which a radio signal is appearing. The visual duty ratio of a signal in visual space may differ from the actual duty ratio in data space. Sim\_CD cannot precisely reflect such differences due to the usage of aggregated CSCP locations. In \autoref{fig:figure7}(a), the visual duty ratio of the signal in Cells 3 and 4 seems large, but the actual duty ratio in Slices 3 and 4 is small. Thus, Dif\_DR is defined to measure such differences.}
\par{The actual duty ratio of a signal can be calculated directly in data space, while the visual duty ratio cannot be directly calculated in visual space. To solve this problem, we thus use the area equivalence principle (\autoref{fig:figure6}(b)) to obtain the visual duty ratios of 10 classes of CSCP distributions in visual space. That is, the ratio of the colored area to the entire area in a cell is used to represent the visual duty ratio of the cell (\autoref{fig:figure6}(c)). Thus, given $n$ slices and $n$ cells, Dif\_DR is formulated as:}
\begin{equation}
	\text{Dif\_DR} = \left| {actual\_DR - \frac{1}{n}\sum\limits_{i = 1}^n {visual\_D{R_i}} } \right|,
	\nonumber
\end{equation}
where $actual\_DR$ represents the actual duty ratio of a signal in data space, which is calculated using the number of logic “1”s divided by the entire time span $T$; and $visual\_DR_i$, which is the visual duty ratio of the $i$th cell. A value approaching 0 indicates a “perfect” result.
\vspace{5pt}
\par{\textbf{(3) The coefficient of variation of time spans of slices (CV\_TS)}}
\vspace{5pt}
\par{In visual design, the size of each cell on the time axis is equal in visual space. If slices in data space have largely differentiated time spans, the visual CSCP distribution and duty ratio presented in visual space may be distorted. Therefore, CV\_TS is defined to measure the dispersion of slices’ time spans. The coefficient of variation is used instead of standard deviation for the calculation because the means of the time spans of the slices in two segmentation results may be different. Given $n$ slices, CV\_TS is formulated as:}
\begin{equation}
	\text{CV\_TS} = \frac{{\sqrt {\frac{1}{n}\sum\nolimits_{i = 1}^n {{{\left( {{t_i} - \mathop t\limits^\_ } \right)}^2}} } }}{{\mathop t\limits^\_ }},
	\nonumber
\end{equation}
where $t_i$ represents the time span of the $i$th slice, and $\overline{t}$  represents the mean of time spans of all slices. A small value indicates a good result.
\vspace{5pt}
\par{\textbf{(4) Loss function design based on the above three metrics}}
\vspace{5pt}
\par{Using the three metrics, the loss function is defined as:}
\begin{equation}
	\begin{array}{l}
		\mathop {\min }\limits_x F(x) = {w_1} \cdot \text{Sim\_CD} + {w_2} \cdot \text{Dif\_DR} + {w_3} \cdot \text{CV\_TS},\\
		{\kern 1pt} {\kern 1pt} {\kern 1pt} {\kern 1pt}{\kern 1pt}{\kern 1pt}{\kern 1pt}{\kern 1pt} {\kern 1pt} {\kern 1pt} {\kern 1pt} {\kern 1pt} {\kern 1pt} {\kern 1pt} {\kern 1pt} {\kern 1pt} {\kern 1pt} {\kern 1pt} {\kern 1pt} {\kern 1pt} {\kern 1pt} {\kern 1pt} {\kern 1pt} {\kern 1pt} {\kern 1pt} {\kern 1pt} {\kern 1pt} {\kern 1pt} {\kern 1pt} {\kern 1pt} {\kern 1pt} {\kern 1pt} {\kern 1pt} {\kern 1pt} {\kern 1pt} {\kern 1pt} {\kern 1pt} {\kern 1pt} {\kern 1pt} {\kern 1pt} {\kern 1pt} {\kern 1pt} {\kern 1pt} {\kern 1pt} {\kern 1pt} {\kern 1pt} {\kern 1pt} {\kern 1pt} {\kern 1pt} {\kern 1pt} {\kern 1pt} {\kern 1pt} {\kern 1pt} {w_1},{w_2},{w_3} \in (0,1]{\kern 1pt} {\rm{ }}{\kern 1pt}{\kern 1pt}\text{and}{\rm{ }}{\kern 1pt}{\kern 1pt} \sum\nolimits_{i = 1}^3 {{w_i} = 1}, 
		\end{array}
		\nonumber
\end{equation}
where $w_i$ is a weight coefficient. The three weights are equal by default but can be adjusted on-demand. For example, we recommend increasing the weight of Sim\_CD for the accurate identification of the beginning or ending time of a communication. We also recommend increasing the weight of Dif\_DR to facilitate the exploration of periodic communication patterns of a signal. A small loss indicates a good time segmentation result. This loss function is an expression of the optimization goal and is used in the algorithm design.

%6.2
\subsection{Algorithm Design}
\label{sec:sec6.2}
To segment a binary sequence of communication states of a signal into time slices, a new time segmentation algorithm, i.e., BSSVA, is designed by making use of a two-strategy and two-stage optimization. Although five time segmentation strategies existed (Section \ref{sec:sec2.3}), no single strategy achieved satisfactory segmentation results in our pilot tests. Interestingly, combining the FP and EL strategies showed potential for two reasons. First, CSCPs are natural feature points for applying FP, but FP has a high time cost and cannot control for the dispersion of time spans of slices. Second, EL can rapidly divide a sequence into slices with equal time spans but ignores CSCPs. Moreover, a satisfactory segmentation result relies on identifying two keys, namely, an optimal number of segments $n$ and $n$-1 optimal dividing points. Our approach adopts a two-stage optimization to identify the two keys in two stages in accordance with the two-strategy combination idea.
\vspace{3pt}
\par{\textbf{STAGE I. Obtain an optimal \emph{n} using an EL strategy.}}
\vspace{3pt}
\par{This stage uses an EL strategy to obtain an optimal $n$ and initial $n-1$ dividing points. Given a sequence and $n$ ranging from \emph{N$_{min}$} to \emph{N$_{max}$} ($n$ $\in$ $n$$^+$), $n$ is traversed from \emph{N$_{min}$} to \emph{N$_{max}$}. On each traversal trial, the sequence is evenly divided by $n-1$ tentative dividing points. Notably, we need to move each tentative dividing point to the nearest integral position on the time axis if the entire period is not divisible by $n$ because the sequence has an interval of 1 second. After traversing, many groups of tentative dividing points are obtained. Each group can generate $n$ tentative slices in data space and $n$ tentative cells in visual space. We can thus calculate the loss of each group by using the loss function from Section \ref{sec:sec6.1}. The group with the smallest loss contributes an optimal $n$ with $n-1$ initial dividing points for the next stage.}
\par{This stage takes minimal time to obtain an optimal $n$ because the time cost of EL is negligible in each traversal step and the range of $n$ is limited in [30, 50] from Section \ref{sec:sec5.2}. This stage obtains initial slices with an equal time span, which is ideal for the CV\_TS metric in the optimization goal. This stage also produces initial dividing points to reduce the searching space of optimal dividing points in the next stage.}
\vspace{3.5pt}
\par{\textbf{STAGE II. Obtain optimal dividing points using an FP strategy.}}
\vspace{3.5pt}

\par
\begin{figure}[h]
	\centering
	\vspace{0.2cm} 	
	\setlength{\abovecaptionskip}{2pt}   %调整图片标题与图距离	
	\setlength{\belowcaptionskip}{-0.5cm}   %调整图片标题与下文距离
	\includegraphics[width=\linewidth]{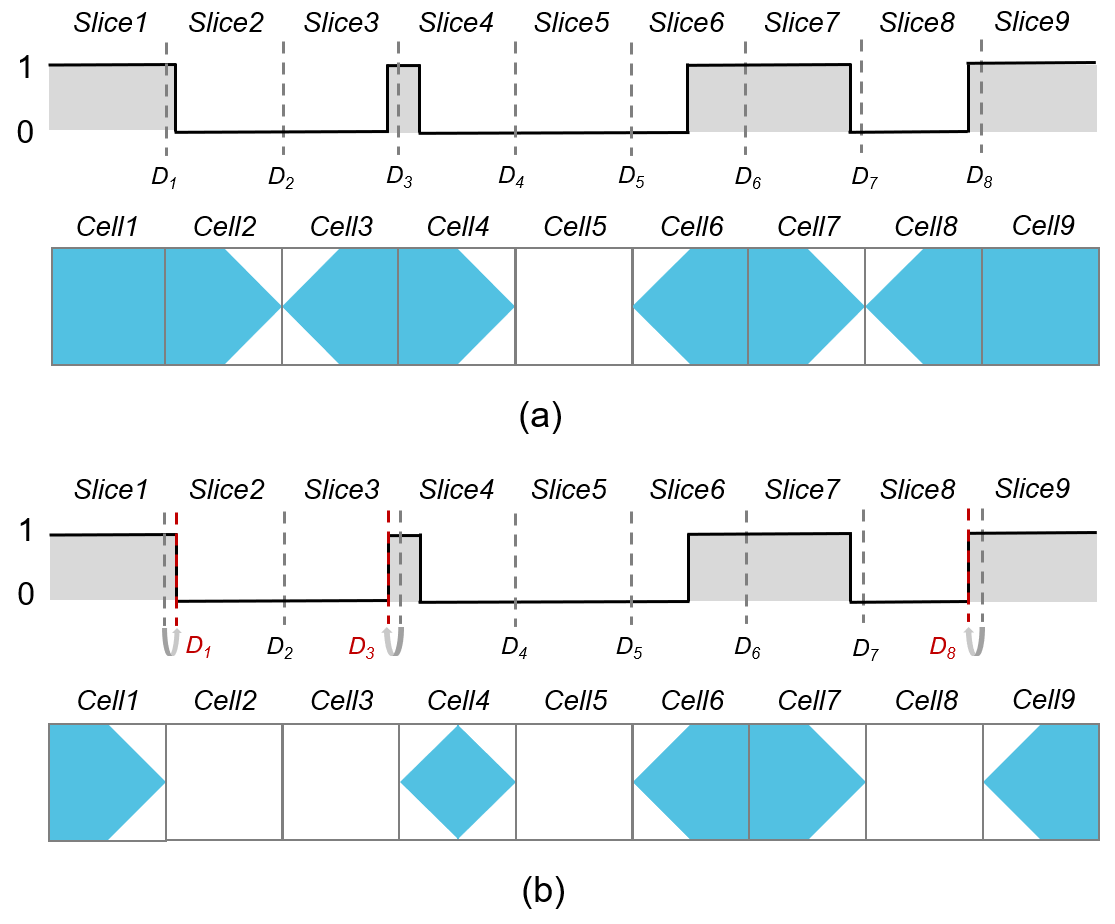}
	\caption{Demonstrations of the two stages of the proposed time segmentation algorithm. (a) Using an EL strategy to divide a binary sequence into 9 slices initially and generate a visual abstraction result in 9 cells. (b) Using an FP strategy to adjust the locations of some initial dividing points on the time axis to generate an improved visual abstraction result.}
	\label{fig:figure7}
\end{figure}

{If the initial dividing points are used to generate visualization, the actual CSCP distribution and duty ratio of a signal in data space may be distorted in visual space. Taking Slices 1$ - $4 and Cells 1$ - $4 in \autoref{fig:figure7}(a) as examples, a long distance exists between the falling edge in Slice2 and the rising edge in Slice3, but the distance is perceived as short in Cells 2 and 3. The duty ratio in Slices 3 and 4 is small but perceived as large in Cells 3 and 4. To prevent such distortions, we need to adjust the locations of initial dividing points. \autoref{fig:figure7}(b) shows an adjustment result, in which the visual CSCP distribution and duty ratio in visual space are perceived as very close to the actual ones in data space.}
\par{We propose two optimization rules and an optimization process to obtain $n-1$ optimal dividing points. The first rule is to restrict the adjustment range such that the location of an initial dividing point can only be adjusted within the time span constructed by its two neighboring time slices. This rule can reduce the searching space and prevent the rapid increase in the dispersion of time spans of slices. The second rule is to generate candidate locations for an initial dividing point. For an initial dividing point \emph{D$_i$} and its two neighboring slices \emph{Slice$_i$} and \emph{Slice}$_\emph{i+1}$, at most two candidate locations exist on the time axis for adjustment, namely, the location of the nearest rising edge in \emph{Slice$_i$} and the location of the nearest falling edge in \emph{Slice}$_\emph{i+1}$. Notably, the location of the nearest falling edge in \emph{Slice$_i$} cannot be a candidate of \emph{D$_i$} due to the belonging principle (\autoref{fig:figure4}(a)), so it is for the nearest rising edge in \emph{Slice}$_\emph{i+1}$. For example, \emph{D}$_\emph{1}$ in \autoref{fig:figure7}(a) has one candidate location, namely, the location of the falling edge in \emph{Slice}$_\emph{2}$. \emph{D}$_\emph{2}$ has no candidate locations, because the edge in \emph{Slice}$_\emph{2}$ is not a rising edge and the edge in \emph{Slice}$_\emph{3}$ is not a falling edge. \emph{D}$_\emph{6}$ has two candidates, namely, the location of the rising edge in \emph{Slice}$_\emph{6}$ and the location of the falling edge in \emph{Slice}$_\emph{7}$.}
\par{The two rules prepare at most $2 \cdot (n-1)$ candidate locations on the basis of the $n-1$ initial dividing points. Our optimization process adopts a gradient descent method with four main steps. (1) Candidate locations are traversed. In each traversal trial, the loss of moving the corresponding initial dividing point to the examining candidate location is calculated. After traversal, we can obtain at most $2\cdot (n-1)$ losses based on the loss function from Section \ref{sec:sec6.1}. (2) The initial dividing point and the candidate location that produce the minimum loss are picked out, and the initial dividing point is moved to the candidate location and marked to not participate in the next iteration. (3) Steps 1 and 2 are repeated until the minimum loss obtained in the current iteration is no longer smaller than the minimum loss in the previous iteration. Taking \autoref{fig:figure7}(b) as an example, the optimization process goes through four iterations. In the first three iterations, \emph{D}$_\emph{1}$, \emph{D}$_\emph{8}$, and \emph{D}$_\emph{3}$ are moved to the locations of the falling edge in Slice2, the rising edge in Slice8, and the rising edge in Slice3, respectively. In the fourth iteration, the optimization process ends because the minimum loss (0.07) is not smaller than the minimum loss (0.06) obtained in the third iteration.}

\begin{table*}[h]
	\begin{center}
		\begin{spacing}{1.2}
			\setlength{\belowcaptionskip}{-0.5cm}
			
		\caption{Results of the algorithm experiment. Bold indicates the best indicator value among the algorithms under the same data conditions.}
		\label{tab:table1}
		\vspace{-0.1cm}

		\setlength{\tabcolsep}{4.3mm}{ %设置单元格的长度 原始的是4.8
		\renewcommand{\arraystretch}{1.25} % default is 1.0
		\begin{tabular}{c c c c c c c c c}
				\Xhline{0.8pt}
				\multirow{2}{*}{Indicator} & \multicolumn{2}{c}{Data} & \multicolumn{6}{c}{Algorithms}\\
				\cline{2-9}
				& Time Span & Complexity & BSSVA (Ours) & EL & SW & BU & TD & FP\\
				\cline{1-9}
				\multirow{4}{*}{Avg loss} & \multirow{2}{*}{A week} & Moderate &\textbf{0.064}&0.076&0.071&0.143&0.139&0.488\\
				\cline{3-9}
				& & High&\textbf{0.114}&0.152&0.128&0.191&0.197&0.315\\
				\cline{2-9}
				&\multirow{2}{*}{A month}&Moderate&\textbf{0.166}&0.181&0.176&0.210&0.225&0.269\\
				\cline{3-9}
				&& High &\textbf{0.176}&0.188&0.185&0.247&0.258&0.270\\
				\cline{1-9}
				\multirow{4}{*}{Avg time (\emph{s})}&\multirow{2}{*}{A week}&Moderate&0.363&\textbf{0.025}&0.342&35.344&4.065&0.177\\
				\cline{3-9}
				&&High&1.361&\textbf{0.040}&0.951&41.174&5.666&1.103\\
				\cline{2-9}
				&\multirow{2}{*}{A month}&Moderate&1.787&\textbf{0.059}&1.254&47.709&7.729&7.239\\
				\cline{3-9}
				&&High&3.071&\textbf{0.090}&1.632&58.168&11.215&18.122\\
				\Xhline{0.8pt}
			\end{tabular}
			\vspace{-1cm}
		}
		
		\end{spacing}
	\end{center}
\end{table*}
% 7
\section{Evaluation}
\label{sec:sec7}

%7.1
\begin{figure*}[bp]
	\centering
	%\vspace{-0.6cm}  %调整图片与上文的垂直距离	
	\setlength{\abovecaptionskip}{1pt}   %调整图片标题与图距离	
	\setlength{\belowcaptionskip}{0cm}   %调整图片标题与下文距离
	\includegraphics[width=\linewidth]{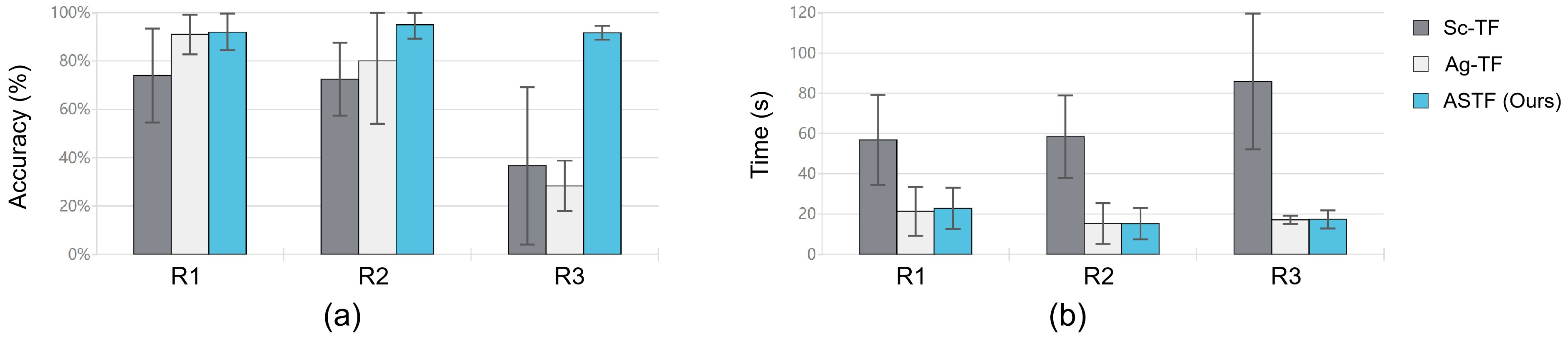}
	\caption{Objective results of the user study in (a) mean accuracy and (b) mean completion time by requirement. Error bars indicate standard errors.}
	\label{fig:figure8}
\end{figure*}
\subsection{Algorithm Performance Experiment}
\label{sec:sec7.1}
An experiment was conducted to verify the performance of the proposed BSSVA algorithm compared to the EL, SW, BU, TD, and FP reference algorithms (Section \ref{sec:sec2.3}). Two prerequisites were set to guarantee experimental fairness. The first was that all algorithms used the same loss function from Section \ref{sec:sec6.1} as the optimization goal. The second was that the number of segments was limited in the same range (i.e., [30, 50]). To keep the two prerequisites, we adaptively modified the reference algorithms. Taking BU as an example, the initial number of small segments was set to 50$\times$10 to reduce the time cost of BU, the loss function helped in determining an optimal segment pair for each merging, and the merger stopped when the loss reached the minimum value in the range of [30, 50]. The supplementary material for this paper details the reference algorithm modifications.
\par{The binary sequences of communication states of 32 radio signals were selected as the experimental data. They were divided into four groups based on time span (i.e., a week or a month) and complexity (i.e., moderate or high). Moderate- and high-complexity signals had an average of 5 and 10 CSCPs per day, respectively. Signals with few CSCPs were not used because they were too simple for the algorithms. Each algorithm ran 10 times on each signal. The average loss and average running time were computed as metrics (\autoref{tab:table1}). Significance analyses with Friedman tests were conducted to examine the differences ($p \ \textless \ 0.05$) among the algorithms in the two metrics. The experiment was conducted on a desktop with a 3.0 GHz Intel i7 CPU, 16 GB memory, and 2K resolution display.}
\par{In terms of average loss, significant differences were found. BSSVA performed significantly better than the other algorithms. SW obtained relatively small losses because we applied a two-stage optimization in the SW modification (see the supplementary material for details). TD and BU had relatively large losses because they could not precisely adjust the locations of the dividing points. The data size and complexity did not influence these results. With regard to the three metrics (results by metric are detailed in the supplementary material), EL performed well in CV\_TS but poorly in Sim\_CD because EL overlooked CSCP locations. FP performed poorly in CV\_TS but well in Sim\_CD, especially for signals with numerous CSCPs, possibly because FP determined dividing points largely based on CSCP locations. These two results supported our two-strategy combination idea for the BSSVA design. SW performed well in CV\_TS but not ideally in Sim\_CD and Dif\_DR due to local optimizations.}
\par{In terms of average time, significant differences were found. EL performed significantly better than the other algorithms. SW and BSSVA had relatively small time costs, but TD and BU exhibited relatively large time costs because SW had a ``one-pass" feature, BSSVA took a small number of iterations (less than the number of slices obtained by EL), while TD and BU required a large number of iterations. FP presented a moderate time performance since its number of iterations was mainly influenced by the number of CSCPs of a signal. Moreover, the influence of the increase in the number of CSCPs on the time costs was greater than that of the increase in the time spans.}

%7.2
\subsection{User Study}
\label{sec:sec7.2}
A user study was conducted to evaluate the effectiveness of the proposed ASTF diagram in long-term radio signal analyses. The scrolling TF diagram (\autoref{fig:figure2}(b)) and aggregated TF diagram (\autoref{fig:figure2}(c)) were selected as references and named as Sc-TF and Ag-TF, respectively. Thirty participants were recruited to participate in the study (i.e., 9 females and 21 males, aged 19$ - $25 years; median age: 21; normal or corrected-to-normal visions). They were graduate students majoring in communication engineering and familiar with using TF diagrams. The participants were evenly divided into three groups, and each group used one of the three diagrams. The user study was conducted on a desktop with a 23.8-inch 1920$\times$1080 display.
\par{During the study, the participants were asked to answer 18 objective questions that required specific information corresponding to one of the analytical requirements (R1$ - $R3 in Section \ref{sec:sec3}) by observing a diagram. For example, “How many radio signals can be found?”, “How many discontinuities does the signal occur?”, “Which of the following four descriptions can reflect the time-varying patterns of the signal?” and so on. Thirty-six datasets with different time spans (i.e., 1 day, 3 days, and 1 week) were prepared. The participants needed to complete two trials for each question with two datasets. After accomplishing their tasks, the participants were asked whether the examined diagram was useful. They rated their overall feelings by requirement using a five-point Likert scale ranging from 1 (strongly disagree) to 5 (strongly agree) and then provided any thoughts they had. The full list of objective questions is provided in the supplementary material.}
\par{The answer and completion time of each participant for each trial were recorded and the mean accuracy and mean time by requirement (\autoref{fig:figure8}) and question (see the supplementary material) were computed as metrics. Significance analyses with Friedman tests were conducted to examine the differences ($p \ \textless \ 0.05$) among the groups in the two metrics and three requirements.}
\par{In terms of accuracy (\autoref{fig:figure8}(a)), no significant differences were found between the Ag-TF and ASTF groups in R1 and R2. The Ag-TF group (91\%) achieved an accuracy close to that of the ASTF group (92\%) in R1, probably because Ag-TF and ASTF had the same basic layout that could explicitly present the number of signals and their basic characteristics. The Ag-TF group had a lower accuracy (80\%) than the ASTF group (95\%) in R2 because some short discontinuities were lost after being aggregated in Ag-TF. The Sc-TF group performed worst in R1 and R2, likely because Sc-TF cannot directly provide an overall picture, and the participants had to scroll to search for the acquired information. Significant differences were found in R3. The ASTF group significantly performed better than the other groups in R3. The Sc-TF group (36\%) outperformed the Ag-TF group (28\%) in R3, possibly because a few anomalies can be found by observing unprocessed raw spectrum data in Sc-TF, even though anomalies were not explicitly encoded in Sc-TF and Ag-TF.} 
\par{In terms of time (\autoref{fig:figure8}(b)), the Sc-TF group took significantly longer to complete R1$ - $R3 due to inevitable scrolling operations when using Sc-TF. No significant differences were found between the ASTF and Ag-TF groups in R1$ - $R3. Most participants in the two groups completed a question in 20 seconds. Moreover, the ASTF group was slightly slower than the Ag-TF group in R1, probably because some participants were not familiar with the visual encodings in ASTF.}

\begin{figure}[h]
	\centering
	%\vspace{-0.6cm}  %调整图片与上文的垂直距离	
	\setlength{\abovecaptionskip}{-2pt}   %调整图片标题与图距离	
	\setlength{\belowcaptionskip}{-0.2cm}   %调整图片标题与下文距离
	\includegraphics[width=\linewidth]{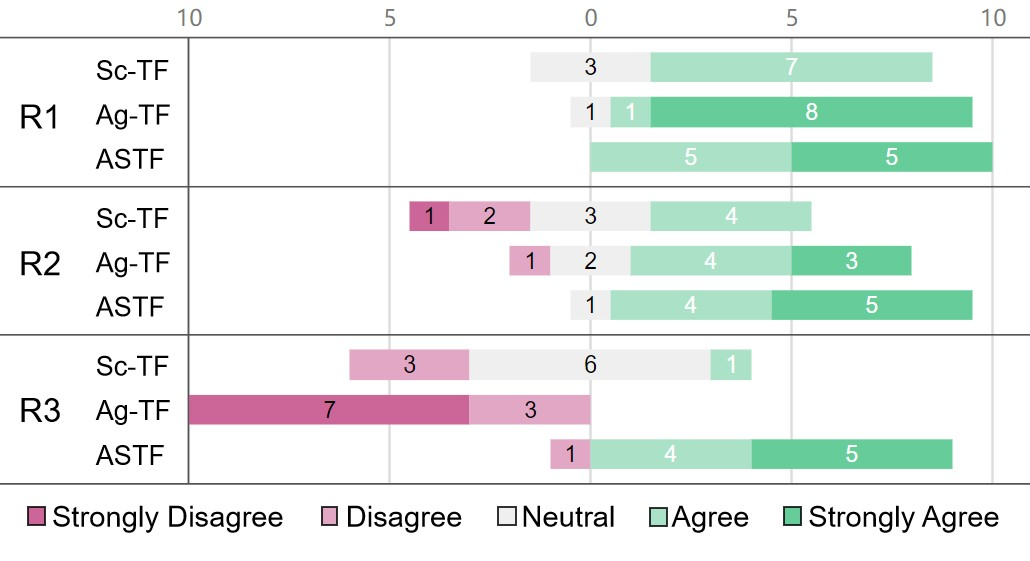}
	\caption{Subjective rating results of the participants in the user study.}
	\label{fig:figure9}
\end{figure}
\par{The subjective ratings of the participants are shown in \autoref{fig:figure9}. The rating medians are as follows. R1: Sc-TF (\emph{m} = 4), Ag-TF (\emph{m} = 5), and ASTF (\emph{m} = 4.5); R2: Sc-TF (\emph{m} = 3), Ag-TF (\emph{m} = 4), and ASTF (\emph{m} = 4.5); R3: Sc-TF (\emph{m} = 3), Ag-TF (\emph{m} = 1), and ASTF (\emph{m} = 4.5). Ag-TF obtained the highest rating median in R1. Several participants commented, “\emph{Ag-TF’s visualization is simple and effective in presenting signal distribution and characteristics.}” Sc-TF obtained the lowest rating medians in R1 and R2. Some participants stated, “\emph{It is exhausting to use Sc-TF. I have to scroll the diagram many times and some details are easily overlooked. Importantly, the previously noticed information is often forgotten after scrolling.}” ASTF obtained the highest rating medians in R2 and R3. Participants made comments such as, “\emph{ASTF is informative and helpful in answering all questions}”, “\emph{It seems [there are] no anomalies in Ag-TF}”, “\emph{Some short discontinuities in the ground truth cannot be seen in Ag-TF. Some short discontinuities are just light white lines, which are indistinguishable,}” and “\emph{It would be great to connect a Sc-TF with an ASTF to form an overview-detail mode}”.}

%8
\section{Discussion}
\label{sec:sec8}

%8.1
\subsection{Limitations}
\label{sec:sec8.1}
ASTF diagrams address the scalability of TF diagrams in terms of time but not frequency. The frequency range that can be utilized by human beings is extremely wide, but raw radio spectrum data generally have narrow bands limited by the spectrum sensing capability of mainstream equipment (e.g., 36 MHz or 72 MHz). Even so, wide-band spectrum data can be obtained through complicated post-processing. Therefore, the scalability for frequency is important and challenging.
\par{ASTF diagrams may encounter three unique problems in practice. First, sparse and dense areas may co-exist in ASTF when signals are unevenly distributed on the frequency. This problem could be solved by providing interval adjustment functions or zooming interactions. Second, the height-width ratio of a cell in visual space could be small when representing a signal with a wide bandwidth. A potential solution could be to set a minimum height-width ratio of a cell in visual space. Lastly, the number of slices of multiple signals may be slightly different if segmenting solely on signals. Generally, such differences have little impact. Finding an optimal $n$ on all signals could avoid this problem.}
\par{The BSSVA algorithm has three limitations. First, the algorithm can not be directly applied to streaming data due to its data pre-processing mode. A potential solution could be to adopt an incremental data processing mode \cite{S1,S2}. Second, the algorithm’s result may not be globally optimal. Two alternatives possibly achieve smaller losses than the algorithm. The one is to find an optimal $n$ in STAGE II instead of STAGE I. The other is to adopt a genetic algorithm for time segmentation \cite{S3,S4}. We tested these two alternatives in pilots and the results seemed to show that losses did not decline substantially but time costs increased greatly. Lastly, the time performance can be improved. A feasible solution is to introduce parallel computing techniques \cite{S5}.}

%8.2
\subsection{Implications}
\label{sec:sec8.2}
ASTF diagrams have the potential to become a common visual component in visual analysis systems for radio monitoring and management. An ASTF diagram could act as a superior view in a multi-viewed interface to provide an overview of the time-varying signals in a large time span. Users thus could select a short time span of interest in the ASTF diagram to update a TF diagram to observe details in raw radio spectrum data. Moreover, ASTF diagrams could provide semantic interactions on signals to fulfill advanced analysis tasks. For instance, when comparing the communication behaviors of two signals, users could select the two signals to filter out the other signals in an ASTF diagram and then juxtapose them closely in frequency to facilitate the comparison. Such interactions cannot be directly supported in TF diagrams because radio spectrum data are not formatted in structured signal records. See the supplementary material for demonstrations.
\par{The visual abstraction method and the BSSVA time segmentation algorithm can be applied to visualize long-term binary sequences generated in various domains for time-varying pattern analyses. For example, a sensor that reports whether a car parking spot is or is not occupied at a certain frequency can produce long-term binary sequences. A monitor of digit circuits can also produce long-term binary sequences. Moreover, common time series data can be converted into binary sequences by setting appropriate thresholds on-demand.}

%9
\section{Conclusion}
\label{sec:sec9}
This paper presents a new visual diagram, ASTF, that can depict the time-varying patterns of radio signals to fulfill long-term signal analysis requirements. A visual abstraction method and a time segmentation algorithm are proposed to assist in implementing the new diagram. The new diagram can be combined with a traditional TF diagram to work in a hierarchical way in a visual analysis system for radio signal analyses. The new diagram can also provide semantic interactions on signals in the system to facilitate advanced signal analysis tasks. The proposed visual abstraction method and time segmentation algorithm can be applied for the visualization of other binary sequences.

\section*{Acknowledgments}
The work is supported in part by the National Natural Science Foundation of China (No. 62072470 and 61872388) and the Fundamental Research Funds for the Central Universities of Central South University. ASTF at Github: \href{https://github.com/csuvis/ASTF}{https://github.com/csuvis/ASTF}.
% references
\bibliographystyle{abbrv-doi}
\bibliography{ASTF}

\begin{thebibliography}{10}

\bibitem{laobaizuishuai}
M.~Adnan, M.~Just, and L.~Baillie.
\newblock Investigating time series visualisations to improve the user
  experience.
\newblock In {\em Proceedings of the 2016 CHI Conference on Human Factors in
  Computing Systems}, pp. 5444--5455, 2016.

\bibitem{D3}
M.~Adnan, M.~Just, and L.~Baillie.
\newblock Investigating time series visualisations to improve the user
  experience.
\newblock In {\em Proceedings of the CHI Conference on Human Factors in
  Computing Systems (CHI), New York, NY, USA}, pp. 5444--5455. ACM, 2016.

\bibitem{A16}
W.~Aigner, S.~Miksch, W.~M{\"u}ller, H.~Schumann, and C.~Tominski.
\newblock Visual methods for analyzing time-oriented data.
\newblock {\em IEEE Transactions on Visualization and Computer Graphics},
  14(1):47--60, 2007.

\bibitem{D15}
W.~Aigner, A.~Rind, and S.~Hoffmann.
\newblock Comparative evaluation of an interactive time-series visualization
  that combines quantitative data with qualitative abstractions.
\newblock {\em Computer Graphics Forum}, 31(3pt2):995--1004, 2012.

\bibitem{A7}
A.~Al~Adnani, J.~Duplicy, and L.~Philips.
\newblock Spectrum analyzers today and tomorrow: part 1 towards
  filterbanks-enabled real-time spectrum analysis.
\newblock {\em IEEE Instrumentation \& Measurement Magazine}, 16(5):6--11,
  2013.

\bibitem{A12}
J.~Alexander, A.~Cockburn, S.~Fitchett, C.~Gutwin, and S.~Greenberg.
\newblock Revisiting read wear: analysis, design, and evaluation of a
  footprints scrollbar.
\newblock In {\em Proceedings of the SIGCHI Conference on Human Factors in
  Computing Systems (CHI), New York, NY, USA}, pp. 1665--1674. ACM, 2009.

\bibitem{X1}
S.~{Alnegheimish}, D.~{Liu}, C.~{Sala}, L.~{Berti-Equille}, and
  K.~{Veeramachaneni}.
\newblock Sintel: A machine learning framework to extract insights from
  signals.
\newblock {\em arXiv preprint arXiv:2204.09108}, 2022.

\bibitem{E7}
J.~Bernard, C.~Bors, M.~B{\"o}gl, C.~Eichner, T.~Gschwandtner, S.~Miksch,
  H.~Schumann, and J.~Kohlhammer.
\newblock Combining the automated segmentation and visual analysis of
  multivariate time series.
\newblock In {\em Proceedings of EuroVis Workshop on Visual Analytics (EuroVA),
  Brno, Czech Republic}, pp. 49--53. IEEE, 2018.

\bibitem{D14}
J.~Bernard, E.~Dobermann, M.~B{\"o}gl, M.~R{\"o}hlig, A.~V{\"o}gele, and
  J.~Kohlhammer.
\newblock Visual-interactive segmentation of multivariate time series.
\newblock In {\em Proceedings of EuroVis Workshop on Visual Analytics (EuroVA),
  Groningen, Netherlands}, pp. 31--35. IEEE, 2016.

\bibitem{D18}
M.~Brehmer, B.~Lee, B.~Bach, N.~H. Riche, and T.~Munzner.
\newblock Timelines revisited: A design space and considerations for expressive
  storytelling.
\newblock {\em IEEE Transactions on Visualization and Computer Graphics},
  23(9):2151--2164, 2016.

\bibitem{G1}
O.~Bryngdahl.
\newblock Moir{\'e}: formation and interpretation.
\newblock {\em Journal of the Optical Society of America}, 64(10):1287--1294,
  1974.

\bibitem{C6}
A.~Cantu, T.~Duval, O.~Grisvard, and G.~Coppin.
\newblock Helovis: A helical visualization for sigint analysis using 3d
  immersion.
\newblock In {\em Proceedings of International Conference on Pacific
  Visualization Symposium (PacificVis), Kobe, Japan}, pp. 175--179. IEEE, 2018.

\bibitem{B3}
D.~Chen, J.~Yang, J.~Wu, H.~Tang, and M.~Huang.
\newblock Spectrum occupancy analysis based on radio monitoring network.
\newblock In {\em Proceedings of International Conference on Communications in
  China (ICC), Beijing, China}, pp. 739--744. IEEE, 2012.

\bibitem{B2}
L.~Chen, D.~Shi, and Y.~Gao.
\newblock Research on the simulation system of the complex electromagnetic
  environment.
\newblock In {\em Proceedings of International Conference on Electromagnetic
  Compatibility (EMC), Tokyo, Japan}, pp. 45--48. IEEE, 2014.

\bibitem{E12}
C.-S.~J. Chu.
\newblock Time series segmentation: A sliding window approach.
\newblock {\em Information Sciences}, 85(1-3):147--173, 1995.

\bibitem{E14}
F.~L.~K. Chung, T.-C. Fu, W.~P.~R. Luk, and V.~T.~Y. Ng.
\newblock Flexible time series pattern matching based on perceptually important
  points.
\newblock In {\em Proceedings of Workshop on Learning from Temporal and Spatial
  Data in International Joint Conference on Artificial Intelligence (IJCAI),
  Seattle, WA, USA}, pp. 1--7. Morgan Kaufmann, 2001.

\bibitem{A13}
A.~Cockburn, A.~Karlson, and B.~B. Bederson.
\newblock A review of overview+detail, zooming, and focus+context interfaces.
\newblock {\em ACM Computing Surveys}, 41(1):1--31, 2009.

\bibitem{A6}
L.~Cohen.
\newblock Time-frequency distributions-a review.
\newblock {\em Proceedings of the IEEE}, 77(7):941--981, 1989.

\bibitem{A8}
M.~Cotton, J.~Wepman, J.~Kub, S.~Engelking, Y.~Lo, H.~Ottke, R.~Kaiser,
  D.~Anderson, M.~Souryal, and M.~Ranganathan.
\newblock An overview of the ntia/nist spectrum monitoring pilot program.
\newblock In {\em Proceedings of International Conference on Wireless
  Communications and Networking Conference Workshops (WCNCW), New Orleans, LA,
  USA}, pp. 217--222. IEEE, 2015.

\bibitem{C8}
T.~Crnovrsanin, C.~Muelder, and K.-L. Ma.
\newblock A system for visual analysis of radio signal data.
\newblock In {\em Proceedings of International Conference on Visual Analytics
  Science and Technology (VAST), Paris, France}, pp. 33--42. IEEE, 2014.

\bibitem{F5}
J.~Ding and J.~Cai.
\newblock Two-side coalitional matching approach for joint mimo-noma clustering
  and bs selection in multi-cell mimo-noma systems.
\newblock {\em IEEE Transactions on Wireless Communications}, 19(3):2006--2021,
  2019.

\bibitem{A15}
N.~Elmqvist and J.-D. Fekete.
\newblock Hierarchical aggregation for information visualization: Overview,
  techniques, and design guidelines.
\newblock {\em IEEE Transactions on Visualization and Computer Graphics},
  16(3):439--454, 2009.

\bibitem{S2}
T.~Fujiwara, J.-K. Chou, S.~Shilpika, P.~Xu, L.~Ren, and K.-L. Ma.
\newblock An incremental dimensionality reduction method for visualizing
  streaming multidimensional data.
\newblock {\em IEEE Transactions on Visualization and Computer Graphics},
  26(1):418--428, 2019.

\bibitem{Z4}
J.~Gao, X.~Yi, C.~Zhong, X.~Chen, and Z.~Zhang.
\newblock Deep learning for spectrum sensing.
\newblock {\em IEEE Wireless Communications Letters}, 8(6):1727--1730, 2019.

\bibitem{S4}
K.~M. Hamdia, X.~Zhuang, and T.~Rabczuk.
\newblock An efficient optimization approach for designing machine learning
  models based on genetic algorithm.
\newblock {\em Neural Computing and Applications}, 33(6):1923--1933, 2021.

\bibitem{D19}
M.~C. Hao, U.~Dayal, D.~A. Keim, and T.~Schreck.
\newblock Multi-resolution techniques for visual exploration of large
  time-series data.
\newblock In {\em Proceedings of the Eurographics / IEEE VGTC conference on
  Visualization (EuroVis), Norrköping, Sweden}, pp. 27--34. IEEE, 2007.

\bibitem{D16}
M.~C. Hao, M.~Marwah, H.~Janetzko, U.~Dayal, D.~A. Keim, D.~Patnaik,
  N.~Ramakrishnan, and R.~K. Sharma.
\newblock Visual exploration of frequent patterns in multivariate time series.
\newblock {\em Information Visualization}, 11(1):71--83, 2012.

\bibitem{B8}
A.~H. Hassan, C.~J. Fluke, and D.~G. Barnes.
\newblock Interactive visualization of the largest radioastronomy cubes.
\newblock {\em New Astronomy}, 16(2):100--109, 2011.

\bibitem{S3}
J.~H. Holland.
\newblock Genetic algorithms.
\newblock {\em Scientific American}, 267(1):66--73, 1992.

\bibitem{D6}
W.~Javed and N.~Elmqvist.
\newblock Stack zooming for multi-focus interaction in time-series data
  visualization.
\newblock In {\em Proceedings of International Conference on Pacific
  Visualization Symposium (PacificVis), Taipei, Taiwan, China}, pp. 33--40.
  IEEE, 2010.

\bibitem{A14}
U.~Jugel, Z.~Jerzak, G.~Hackenbroich, and V.~Markl.
\newblock M4: a visualization-oriented time series data aggregation.
\newblock {\em Proceedings of the VLDB Endowment}, 7(10):797--808, 2014.

\bibitem{E4}
E.~Keogh.
\newblock Fast similarity search in the presence of longitudinal scaling in
  time series databases.
\newblock In {\em Proceedings of International Conference on Tools with
  Artificial Intelligence (ICTAI), Newport Beach, CA, USA}, pp. 578--584. IEEE,
  1997.

\bibitem{E9}
E.~Keogh, K.~Chakrabarti, M.~Pazzani, and S.~Mehrotra.
\newblock Dimensionality reduction for fast similarity search in large time
  series databases.
\newblock {\em Knowledge and Information Systems}, 3(3):263--286, 2001.

\bibitem{E11}
E.~Keogh, S.~Chu, D.~Hart, and M.~Pazzani.
\newblock An online algorithm for segmenting time series.
\newblock In {\em Proceedings of International Conference on Data Mining
  (ICDM), San Jose, CA, USA}, pp. 289--296. IEEE, 2001.

\bibitem{C3}
R.~Kincaid.
\newblock Signallens: Focus+context applied to electronic time series.
\newblock {\em IEEE Transactions on Visualization and Computer Graphics},
  16(6):900--907, 2010.

\bibitem{D17}
S.~Ko, S.~Afzal, S.~Walton, Y.~Yang, J.~Chae, A.~Malik, Y.~Jang, M.~Chen, and
  D.~Ebert.
\newblock Analyzing high-dimensional multivariate network links with integrated
  anomaly detection, highlighting and exploration.
\newblock In {\em Proceedings of International Conference on Visual Analytics
  Science and Technology (VAST), Paris, France}, pp. 83--92. IEEE, 2014.

\bibitem{B1}
A.~Kostic and D.~Rancic.
\newblock Radar coverage analysis in virtual gis environment.
\newblock In {\em Proceedings of International Conference on Telecommunications
  in Modern Satellite, Cable and Broadcasting Service (TELSIKS), Nis,
  Yugoslavia}, pp. 721--724. IEEE, 2003.

\bibitem{D13}
R.~Kr{\"u}ger, D.~Thom, M.~W{\"o}rner, H.~Bosch, and T.~Ertl.
\newblock Trajectorylenses--a set-based filtering and exploration technique for
  long-term trajectory data.
\newblock {\em Computer Graphics Forum}, 32(3pt4):451--460, 2013.

\bibitem{D11}
N.~Kumar, V.~N. Lolla, E.~Keogh, S.~Lonardi, C.~A. Ratanamahatana, and L.~Wei.
\newblock Time-series bitmaps: a practical visualization tool for working with
  large time series databases.
\newblock In {\em Proceedings of the SIAM International Conference on Data
  Mining (SDM), Newport Beach, CA, USA}, pp. 531--535. SIAM, 2005.

\bibitem{A11}
H.~Lam.
\newblock A framework of interaction costs in information visualization.
\newblock {\em IEEE Transactions on Visualization and Computer Graphics},
  14(6):1149--1156, 2008.

\bibitem{S5}
J.~K. Li and K.-L. Ma.
\newblock P4: Portable parallel processing pipelines for interactive
  information visualization.
\newblock {\em IEEE Transactions on Visualization and Computer Graphics},
  26(3):1548--1561, 2018.

\bibitem{E1}
W.~Long, Z.~Lu, and L.~Cui.
\newblock Deep learning-based feature engineering for stock price movement
  prediction.
\newblock {\em Knowledge-Based Systems}, 164:163--173, 2019.

\bibitem{C4}
R.~Lopez-Hernandez, D.~Guilmaine, M.~J. McGuffin, and L.~Barford.
\newblock A layer-oriented interface for visualizing time-series data from
  oscilloscopes.
\newblock In {\em Proceedings of International Conference on Pacific
  Visualization Symposium (PacificVis), Taipei, Taiwan, China}, pp. 41--48.
  IEEE, 2010.

\bibitem{E8}
M.~Lovri{\'c}, M.~Milanovi{\'c}, and M.~Stamenkovi{\'c}.
\newblock Algoritmic methods for segmentation of time series: An overview.
\newblock {\em Journal of Contemporary Economic and Business Issues},
  1(1):31--53, 2014.

\bibitem{E5}
K.~B. Pratt and E.~Fink.
\newblock Search for patterns in compressed time series.
\newblock {\em International Journal of Image and Graphics}, 2(01):89--106,
  2002.

\bibitem{D10}
T.~Saito, H.~N. Miyamura, M.~Yamamoto, H.~Saito, Y.~Hoshiya, and T.~Kaseda.
\newblock Two-tone pseudo coloring: Compact visualization for one-dimensional
  data.
\newblock In {\em Proceedings of International Conference on Information
  Visualization (INFOVIS), Minneapolis, MN, USA}, pp. 173--180. IEEE, 2005.

\bibitem{B4}
F.~Salim, M.~Williams, N.~Sony, M.~D. Pena, Y.~Petrov, A.~A. Saad, and B.~Wu.
\newblock Visualization of wireless sensor networks using zigbee's received
  signal strength indicator (rssi) for indoor localization and tracking.
\newblock In {\em Proceedings of International Conference on Pervasive
  Computing and Communication Workshops (PERCOM WORKSHOPS), Budapest, Hungary},
  pp. 575--580. IEEE, 2014.

\bibitem{C5}
N.~Sharakhov, V.~Marojevic, F.~Romano, N.~Polys, and C.~Dietrich.
\newblock Visualizing real-time radio spectrum access with cornet3d.
\newblock In {\em Proceedings of International Conference on 3D Web
  Technologies (Web3D), New York, NY, USA}, pp. 109--116. ACM, 2014.

\bibitem{F4}
C.~Shen, X.~Bao, J.~Tan, S.~Liu, and Z.~Liu.
\newblock Two noise-robust axial scanning multi-image phase retrieval
  algorithms based on pauta criterion and smoothness constraint.
\newblock {\em Optics Express}, 25(14):16235--16249, 2017.

\bibitem{A10}
G.~Shurkhovetskyy, N.~Andrienko, G.~Andrienko, and G.~Fuchs.
\newblock Data abstraction for visualizing large time series.
\newblock {\em Computer Graphics Forum}, 37(1):125--144, 2018.

\bibitem{E13}
Y.-W. Si and J.~Yin.
\newblock Obst-based segmentation approach to financial time series.
\newblock {\em Engineering Applications of Artificial Intelligence},
  26(10):2581--2596, 2013.

\bibitem{B9}
Z.~Song, Y.~Zhang, H.~Lv, P.~Chen, and C.~Tang.
\newblock Electromagnetic situation generation algorithm based on information
  geometry.
\newblock {\em Telecommunication Systems}, 77(1):171--187, 2021.

\bibitem{S1}
Y.~Tanahashi, C.-H. Hsueh, and K.-L. Ma.
\newblock An efficient framework for generating storyline visualizations from
  streaming data.
\newblock {\em IEEE Transactions on Visualization and Computer Graphics},
  21(6):730--742, 2015.

\bibitem{Z1}
H.~van~de Wetering, N.~Klaassen, and M.~Burch.
\newblock Space-reclaiming icicle plots.
\newblock In {\em Proceedings of International Conference on Pacific
  Visualization Symposium (PacificVis), Tianjin, China}, pp. 121--130. IEEE,
  2020.

\bibitem{B10}
R.~Vijgen.
\newblock Architecture of radio.
\newblock \url{http://www.architectureofradio.com}, 2019.
\newblock Accessed on 14 January 2019.

\bibitem{D5}
J.~Walker, R.~Borgo, and M.~W. Jones.
\newblock Timenotes: a study on effective chart visualization and interaction
  techniques for time-series data.
\newblock {\em IEEE Transactions on Visualization and Computer Graphics},
  22(1):549--558, 2015.

\bibitem{E3}
Y.~Wang, P.~Wang, J.~Pei, W.~Wang, and S.~Huang.
\newblock A data-adaptive and dynamic segmentation index for whole matching on
  time series.
\newblock {\em Proceedings of the VLDB Endowment}, 6(10):793--804, 2013.

\bibitem{Z3}
Wiki.
\newblock Moire\_pattern.
\newblock \url{https://en.wikipedia.org/wiki/Moire\_pattern}, 2022.
\newblock Accessed on 9 January 2022.

\bibitem{A1}
Wiki.
\newblock Radio communication.
\newblock \url{https://en.wikipedia.org/wiki/Radio\#Radio\_communication},
  2022.
\newblock Accessed on 28 February 2022.

\bibitem{Z2}
L.~Woodburn, Y.~Yang, and K.~Marriott.
\newblock Interactive visualisation of hierarchical quantitative data: an
  evaluation.
\newblock In {\em Proceedings of International Conference on Visualization
  Conference (VIS), Vancouver, BC, Canada}, pp. 96--100. IEEE, 2019.

\bibitem{E2}
E.~Zdravevski, P.~Lameski, V.~Trajkovik, A.~Kulakov, I.~Chorbev, R.~Goleva,
  N.~Pombo, and N.~Garcia.
\newblock Improving activity recognition accuracy in ambient-assisted living
  systems by automated feature engineering.
\newblock {\em IEEE Access}, 5:5262--5280, 2017.

\bibitem{Z5}
J.~Zhang, F.~Wang, O.~A. Dobre, and Z.~Zhong.
\newblock Specific emitter identification via hilbert--huang transform in
  single-hop and relaying scenarios.
\newblock {\em IEEE Transactions on Information Forensics and Security},
  11(6):1192--1205, 2016.

\bibitem{E15}
Z.~Zhang, J.~Jiang, and H.~Wang.
\newblock A new segmentation algorithm to stock time series based on pip
  approach.
\newblock In {\em Proceedings of International Conference on Wireless
  Communications, Networking and Mobile Computing (WiCOM), Shanghai, China},
  pp. 5609--5612. IEEE, 2007.

\bibitem{D4}
J.~Zhao, F.~Chevalier, E.~Pietriga, and R.~Balakrishnan.
\newblock Exploratory analysis of time-series with chronolenses.
\newblock {\em IEEE Transactions on Visualization and Computer Graphics},
  17(12):2422--2431, 2011.

\bibitem{X2}
Y.~Zhao, H.~Jiang, Y.~Qin, H.~Xie, Y.~Wu, S.~Liu, Z.~Zhou, J.~Xia, F.~Zhou,
  et~al.
\newblock Preserving minority structures in graph sampling.
\newblock {\em IEEE Transactions on Visualization and Computer Graphics},
  27(2):1698--1708, 2020.

\bibitem{A2}
Y.~Zhao, X.~Luo, X.~Lin, H.~Wang, X.~Kui, F.~Zhou, J.~Wang, Y.~Chen, and
  W.~Chen.
\newblock Visual analytics for electromagnetic situation awareness in radio
  monitoring and management.
\newblock {\em IEEE Transactions on Visualization and Computer Graphics},
  26(1):590--600, 2019.

\bibitem{X3}
Y.~Zhao, L.~Wang, S.~Li, F.~Zhou, X.~Lin, Q.~Lu, and L.~Ren.
\newblock A visual analysis approach for understanding durability test data of
  automotive products.
\newblock {\em ACM Transactions on Intelligent Systems and Technology},
  10(6):1--23, 2019.

\bibitem{B12}
G.~Zhongwei and L.~Bin.
\newblock Visualization of radar electromagnetic waves with jamming.
\newblock {\em Procedia Computer Science}, 107:660--666, 2017.

\end{thebibliography}
\end{document}